\RequirePackage{fix-cm}
\documentclass[11pt]{article}
\usepackage{graphicx}
\usepackage{here}
\usepackage[german,english]{babel}
\usepackage{amsmath}
\usepackage{pdfpages}			
\usepackage[percent]{overpic} 		
\usepackage{rotating}			
\usepackage{caption}			
\begin{document}
%
%
%
\newcommand{\va}{{\bf a}} \newcommand{\vA}{{\bf A}}
\newcommand{\vb}{{\bf b}} \newcommand{\vB}{{\bf B}}
\newcommand{\vc}{{\bf c}} \newcommand{\vC}{{\bf C}}
\newcommand{\vd}{{\bf d}} \newcommand{\vD}{{\bf D}}
\newcommand{\ve}{{\bf e}} \newcommand{\vE}{{\bf E}}
\newcommand{\vf}{{\bf f}} \newcommand{\vF}{{\bf F}}
\newcommand{\vg}{{\bf g}} \newcommand{\vG}{{\bf G}}
\newcommand{\vh}{{\bf h}} \newcommand{\vH}{{\bf H}}
\newcommand{\vi}{{\bf i}} \newcommand{\vI}{{\bf I}}
\newcommand{\vj}{{\bf j}} \newcommand{\vJ}{{\bf J}}
\newcommand{\vk}{{\bf k}} \newcommand{\vK}{{\bf K}}
\newcommand{\vl}{{\bf l}} \newcommand{\vL}{{\bf L}}
\newcommand{\vm}{{\bf m}} \newcommand{\vM}{{\bf M}}
\newcommand{\vn}{{\bf n}} \newcommand{\vN}{{\bf N}}
\newcommand{\vo}{{\bf o}} \newcommand{\vO}{{\bf O}}
\newcommand{\vp}{{\bf p}} \newcommand{\vP}{{\bf P}}
\newcommand{\vq}{{\bf q}} \newcommand{\vQ}{{\bf Q}}
\newcommand{\vr}{{\bf r}} \newcommand{\vR}{{\bf R}}
\newcommand{\vs}{{\bf s}} \newcommand{\vS}{{\bf S}}
\newcommand{\vt}{{\bf t}} \newcommand{\vT}{{\bf T}}
\newcommand{\vu}{{\bf u}} \newcommand{\vU}{{\bf U}}
\newcommand{\vv}{{\bf v}} \newcommand{\vV}{{\bf V}}
\newcommand{\vw}{{\bf w}} \newcommand{\vW}{{\bf W}}
\newcommand{\vx}{{\bf x}} \newcommand{\vX}{{\bf X}}
\newcommand{\vy}{{\bf y}} \newcommand{\vY}{{\bf Y}}
\newcommand{\vz}{{\bf z}} \newcommand{\vZ}{{\bf Z}}


\newcommand{\vxi}{{\mbox{\boldmath$\xi$}}}
\newcommand{\vlambda}{\mbox{\boldmath$\lambda$}} \newcommand{\vLambda}{\mbox{\boldmath$\Lambda$}}
\newcommand{\vDelta}{\mbox{\boldmath$\Delta$}}
\newcommand{\vna}{\mbox{\boldmath$\nabla$}}
\newcommand{\vtau}{\mbox{\boldmath$\tau$}}
\newcommand{\vsigma}{\mbox{\boldmath$\sigma$}}
\newcommand{\vepsilon}{\mbox{\boldmath$\varepsilon$}}
\newcommand{\vPsi}{\mbox{\boldmath$\Psi$}}
\newcommand{\vvarphi}{\mbox{\boldmath$\varphi$}}
\newcommand{\vchi}{\mbox{\boldmath$\chi$}}
\newcommand{\valpha}{\mbox{\boldmath$\alpha$}}
\newcommand{\vbeta}{\mbox{\boldmath$\beta$}}
\newcommand{\vgamma}{\mbox{\boldmath$\gamma$}}
\newcommand{\vnu}{\mbox{\boldmath$\nu$}}
%
%
\newcommand{\bdm}{\begin{displaymath}} \newcommand{\edm}{\end{displaymath}}
\newcommand{\nn}{\nonumber}
\newcommand{\bc}{\begin{center}} \newcommand{\ec}{\end{center}}
\newcommand{\be}{\begin{equation}} \newcommand{\ee}{\end{equation}}
\newcommand{\ra}{{\rm a}} \newcommand{\rb}{{\rm b}}
\newcommand{\e}{{\rm e}} \newcommand{\rc}{{\rm c}} \newcommand{\rh}{{\rm h}}
\newcommand{\kB}{k_{\rm B}}
\newcommand{\kF}{k_{\rm F}} \newcommand{\EF}{E_{\rm F}}
\newcommand{\NF}{N_{\rm F}} \newcommand{\pF}{p_{\rm F}}
\newcommand{\Tc}{T_{\rm c}} \newcommand{\vvF}{v_{\rm F}}
\newcommand{\vPi}{{\bf\Pi}} \newcommand{\muB}{\mu_{\rm B}}
\newcommand{\lela}{\left\langle} \newcommand{\rira}{\right\rangle}
%
%
\title{\bf Two--fluid description of two--band superconductors}
\author{Nikolaj Bittner$^{1,2}$ and Dietrich Einzel$^2$,\\
$^1$ Max--Planck--Institut f\"ur Festk\"orperforschung, \\
Heisenbergstra\ss e 1, D--70569 Stuttgart, FRG \\
$^2$ Walther--Mei\ss ner--Institut f\"ur Tieftemperaturforschung, \\
Bayerische Akademie der Wissenschaften, \\
Walther--Mei\ss ner--Stra\ss e 8, D--85748 Garching, FRG}

\date{\today}
\maketitle
\bc
{\bf Abstract} \\[3ex]
\ec
{\it We present a systematic study of the response properties of two--band (multi--gap) superconductors with spin--singlet
($s$--wave) pairing correlations, which are assumed to be caused by both intraband ($\lambda_{ii}$, $i=1,2$) and interband
($\lambda_{12}$) pairing interactions. In this first of three planned publications we concentrate on the properties of such
superconducting systems in global and local thermodynamic equilibrium, the latter including weak perturbations in the
stationary long--wavelength limit.
The discussion of global thermodynamic equilibrium must include the solution (analytical in the Ginzburg--Landau and the
low temperature limit) of the coupled self--consistency equations for the two energy gaps $\Delta_i(T),\ i=1,2$. These
solutions allow to study non--universal behavior of the two relevant BCS--M\"uhlschlegel parameters, namely the specific heat
discontinuity $\Delta C/C_N$ and the zero temperature gaps $\Delta_i(0)/\pi\kB\Tc,\ i=1,2$.
The discussion of a local equilibrium situation includes the calculation of the supercurrent density as a property of the
condensate, and the calculation of both the specific heat capacity and the spin susceptibility as properties of the gas of
thermal excitations in the spirit of a microscopic two--fluid description. Non--monotonic behavior in the temperature dependences
of the gaps and all these local response functions is predicted to occur particularly for very small values of the
interband pair--coupling constant $\lambda_{12}$.}\\
%
\newpage
%
%
\section{Introduction}
This work is devoted to a comprehensive investigation of general two--band superconductors with Cooper pairs in a relative
spin--singlet state. This investigation is based on the celebrated BCS theory of superconductivity~\cite{BCS1957} and its modifications
due to Bogoliubov~\cite{BOGMETH}, Valatin~\cite{VALATIN1958}, Gor'kov~\cite{GORKOV1959} and Nambu~\cite{NAMBU1960}. \\
\vspace{0.1mm}\\
In a series of previous publications, one of the authors investigated universal, 
so--called BCS--M\"uhl--schlegel~\cite{MUEHLSCHLEGEL1959} parameters in unconventional superconductors~\cite{EINZEL2002},
as well as the possibility of establishing a microscopic two--fluid description applied to conventional~\cite{EINZEL2003A} 
and unconventional~\cite{EINZEL2003B} superconductors. These considerations were limited to superconductors 
in which the paired electrons reside on a single band. Very early after the publication of the BCS theory extensions of the description 
to more than one band have been investigated~\cite{SMANDW1959}. More recently, multi--band superconductors like MgB$_2$~\cite{MGB2}, 
pnictides~\cite{PNICTIDES} and non--centrosymmetric superconductors~\cite{NCSBOOK2012} have attracted
great attention. On the theory side recent calculations of the superfluid density and 
the specific heat~\cite{KOGANETAL2009} as well as a Ginzburg--Landau 
analysis~\cite{MILOSEVICETAL1,MILOSEVICETAL2,MILOSEVICETAL3} of two--band superconductors are worth mentioning. 
It turns out that the theoretical treatment of pnictides requires in particular an extension 
to a three--band description~\cite{MARCIANIETAL}. The purpose of this paper is therefore an extension of the
discussion of both BCS--M\"uhlschlegel parameters and the local response functions and the associated two--fluid description to
two--band (two--gap) superconductors, to begin with. In this paper, which is the first of a series of 
three publications~\cite{BANDE-2-2013,BANDE-3-2013}, we provide a comprehensive discussion of the behavior of general two--band superconductors
first in the global equilibrium and second in the local equilibrium state, respectively. In the first part we consider the general extension
of the weak coupling BCS theory to two bands and two gaps and the consequent (analytical and numerical) solution of the coupled gap
equations. The second part is devoted to the local response of these superconductors to weak perturbations such as a local superflow velocity, a
local temperature change and a local magnetic field. The associated local response functions, namely the supercurrent density (or equivalently,
the magnetic penetration depth), the specific heat and the spin susceptibility, respectively, are detectable experimentally and
therefore allow for a comparison of theory with experiment. Experimental activities often include studies of the electromagnetic response
and the electronic Raman response, which are performed in a regime of frequencies and wavenumbers far beyond the local equilibrium. This
is why we decided to shift the discussion of the kinetic theory of two--band superconductors to a second publication~\cite{BANDE-2-2013}.
This discussion will include aspects as important as the Nambu--Goldstone or gauge mode~\cite{NAMBU2009}, a new massive collective mode
characterizing the phase dynamics of the order parameter, the so--called Leggett--mode~\cite{LEGGETT1966}, the condensate plasma mode and the shift
of the gauge mode to the plasma frequency, sometimes referred to as the  Higgs mechanism~\cite{ANDERSON1963}. 
Finally, a comprehensive treatment of the electronic Raman response, to which Leggett's collective mode as well as a two--peak structure 
similar to non--centrosymmetric superconductors~\cite{KME2009} is found to contribute, will be published in a third paper~\cite{BANDE-3-2013}. \\
\vspace{0.1mm}\\
This paper is organized as follows: In section~\ref{sec:genBCS} we establish our notation by generalizing the BCS Hamilton operator to the case
of a weakly--coupled two--band (multi--gap) superconductor and elucidate its structure in Nambu space. Next we discuss the coupled
self--consistency equations for the two gaps $\Delta_1(T)$ and $\Delta_2(T)$ in the Ginzburg--Landau (GL) and the low temperature limit
and present numerical results for intermediate temperatures. In section~\ref{sec:LocRespFunc} we derive expressions for all relevant 
local response functions for two--gap superconductors, namely the magnetic penetration depth, the specific heat discontinuity and 
the spin susceptibility and discuss numerical results for the temperature dependence of these quantities for various values of 
the intra-- and inter--band pairing interaction parameters. Section~\ref{sec:SumCon} is devoted to our discussion and conclusions.

\section{Two--band superconductors with singlet pairing correlations}
\label{sec:genBCS}
\subsection{BCS--Leggett Hamiltonian}
Let us consider a superconductor, which is characterized by electrons, that occupy two bands $i=1,2$.
The electrons on these bands are assumed to be created by the operators $\hat{c}_{\vk\sigma i}^\dagger$, $i=1,2$. 
In terms of these operators, the BCS--Hamiltonian can be written as \cite{LEGGETT1966}
\begin{eqnarray}
\label{eq:BCSHam}
\hat{H}&=&\sum_{\vk\sigma i}\xi_{\vk\sigma i}\hat{c}_{\vk\sigma i}^\dagger\hat{c}_{\vk\sigma i}
+\underbrace{\sum_{\vk\vp i}\Gamma_{\vk\vp}^{(ii)}
\hat{c}_{\vk\uparrow i}^\dagger\hat{c}_{-\vk\downarrow i}^\dagger\hat{c}_{-\vp\downarrow i}\hat{c}_{\vp\uparrow i}}_{\rm pairing\ on\ band\ i}
+\underbrace{\sum_{\vk\vp}\Gamma_{\vk\vp}^{(12)}\left\{\hat{c}_{\vk\uparrow 1}^\dagger\hat{c}_{-\vk\downarrow 1}^\dagger
\hat{c}_{-\vp\downarrow 2}\hat{c}_{\vp\uparrow 2}+{\rm H.c.}\right\}}_{\rm mixing}\ .
\end{eqnarray}
Here $\xi_{\vk\sigma i}=\epsilon_{\vk\sigma i}-\mu_i$ is the energy measured from the chemical potential $\mu_i$ in the $i$--th band, 
whereas the quantities $\Gamma_{\vk\vp}^{(ij)}$, $i,j=1,2$ represent the pairing interactions. Like in the one--band case, 
we shall denote the Fermion occupation number operators on the bands $i=1,2$
\begin{eqnarray*}
\hat{n}_{\vk\sigma i}=\hat{c}_{\vk\sigma i}^\dagger\hat{c}_{\vk\sigma i}
\ \ ; \ \ \hat{n}_{i}=\frac{1}{V}\sum_{\vk\sigma}\hat{n}_{\vk\sigma i}=\frac{\hat{N}_i}{V}\ .
\end{eqnarray*}
In the same way, we may introduce BCS pair operators for the two--band case:
\begin{eqnarray*}
\hat{g}_{\vk i}=\hat{c}_{-\vk\downarrow i}\hat{c}_{\vk\uparrow i}
\end{eqnarray*}
with the aid of which the Hamiltonian assumes the form
\begin{eqnarray*}
\hat{H}&=&\sum_{\vk\sigma i}\xi_{\vk\sigma i}\hat{n}_{\vk\sigma i}
+\sum_{\vk i}\hat{g}_{\vk i}^{\dagger}\underbrace{\sum_{\vp j}\Gamma_{\vk\vp}^{(ij)}\hat{g}_{\vp j}}_{=\hat{\Delta}_{\vk i}}\ .
\end{eqnarray*}
This form gives rise to the following definition of pair potential operators, as generalized to the two--band case:
\begin{eqnarray}
\hat{\Delta}_{\vk i}=\sum_{\vp j}\Gamma_{\vk\vp}^{(ij)}\hat{g}_{\vp j}
\end{eqnarray}
by means of which the Hamiltonian assumes the compact form:
\begin{eqnarray*}
\hat{H}&=&\sum_{\vk\sigma i}\xi_{\vk\sigma i}\hat{n}_{\vk\sigma i}+\sum_{\vk i}\hat{\Delta}_{\vk i}\hat{g}_{\vk i}^\dagger
\end{eqnarray*}
We have now a convenient starting point for performing a {\it mean field approximation} by introducing 
the Gor'kov pair amplitudes $g_{\vk i}$, $i=1,2$~\cite{GORKOV1959}:
\begin{eqnarray}
\hat{g}_{\vk i}\to{g}_{\vk i}=\lela\hat{c}_{-\vk\downarrow i}\hat{c}_{\vk\uparrow i}\rira\not=0 \ \ \ \ {\rm only\ for}\ \ T\leq\Tc
\end{eqnarray}
from which the following form of the two mean pair potentials $\Delta_{\vk i}=\lela\hat{\Delta}_{\vk i}\rira, i=1,2$, may be deduced:
\begin{eqnarray}
\label{eq:EnGap}
\Delta_{\vk i}&=&\sum_{\vp j}\Gamma_{\vk\vp}^{(ij)}{g}_{\vp j}
\end{eqnarray}
The BCS Hamiltonian reads finally
\begin{eqnarray}
\hat{H}_{\rm BCS}&=&\sum_{\vk\sigma i}\xi_{\vk\sigma i}\hat{n}_{\vk\sigma i}
+\sum_{\vk i}\left\{\Delta_{\vk i}\hat{g}_{\vk i}^{\dagger}+\Delta_{\vk i}^{*}\hat{g}_{\vk i}\right\}\ .
\end{eqnarray}
\subsection{Two--band superconductors in global thermodynamic equilibrium}
Let us proceed by combining the energies $\xi_{\vk i}$ and $\Delta_{\vk i}$ into
energy matrices in Nambu space~\cite{NAMBU1960} appropriate for two--band superconductors (two--band Nambu space) in the form
\begin{eqnarray}
\label{eq:DefEnMat}
\underline{\xi}_{\vk i}^0&\equiv& \left( \begin{array}{cc}
\xi_{\vk i}             &    \Delta_{\vk i} \\
\Delta_{\vk i}^*        &    -\xi_{-\vk i}
\end{array} \right)\ .
\end{eqnarray}
Using the energy matrix $\underline{\xi}_{\vk i}^0$, the BCS Hamilton operator can be written in a way reminiscent
of the normal state
\begin{eqnarray*}
\hat H_{\rm BCS}=\sum_{\vk i}\ \underline{\hat C}^{\dagger}_{\vk i}
\cdot\underline{\xi}_{\vk i}^0\cdot\underline{\hat C}_{\vk i}
\end{eqnarray*}
where we have defined spinor creation and annihilation operators in the two--band Nambu space
\begin{eqnarray*}
\underline{\hat C}_{\vk i}&\equiv&\left( \begin{array}{c}
\hat c_{\vk\uparrow i}   \\
\hat c^{\dagger}_{-\vk\downarrow i} \\
\end{array} \right) \\
\underline{\hat C}^{\dagger}_{\vk i}&\equiv&\left( \begin{array}{cc}
\hat c^{\dagger}_{\vk\uparrow i} &  \hat c_{-\vk\downarrow i}
\end{array} \right)\ .
\end{eqnarray*}
We show next that the energy matrix (\ref{eq:DefEnMat}) can be diagonalized by a Bogoliubov--Valatin matrix~\cite{BOGMETH,VALATIN1958}, 
readily generalized to the two--band case. The latter can be written in the form:
\begin{eqnarray*}
\underline{U}_{\vk i}&\equiv& \left(\begin{array}{cc}
u_{\vk i}          &   v_{\vk i} \\
-v_{\vk i}^*       &  u_{\vk i}
\end{array} \right)
\end{eqnarray*}
with $u_{\vk i}$ and $v_{\vk i}$ denoting the usual BCS coherence factors
\begin{eqnarray*}
u_{\vk i}^2=\frac{1}{2}\left(1+\frac{\xi_{\vk i}}{E_{\vk i}}\right)=1-v_{\vk i}^2  
\end{eqnarray*}
for the two bands $i=1,2$. One can easily demonstrate, that this definition of the Bogoliubov--Valatin matrix leads to 
the following result:
\begin{eqnarray}
\underline{U}_{\vk i}^\dagger\cdot\underline{\xi}_{\vk i}^0\cdot\underline{U}_{\vk i}
\equiv& \left( \begin{array}{cc}
E_{\vk i}       &   0   \\
0               & -E_{\vk i}
\end{array} \right)\ \ ;\ \ E_{\vk i}=\sqrt{\xi_{\vk i}^2+|\Delta_{\vk i}|^2}
\end{eqnarray}
After the Bogoliubov--Valatin diagonalization with
\begin{eqnarray*}
 \underline{\hat C}_{\vk i}&=&\underline{U}_{\vk i}\cdot\underline{\hat\alpha}_{\vk i} \ \ \ ; \ \ \
\underline{\hat\alpha}_\vk=\left( \begin{array}{c}
\hat \alpha_{\vk\uparrow i}\\
\hat\alpha^{\dagger}_{-\vk\downarrow i}
\end{array} \right)
\end{eqnarray*}
the BCS Hamiltonian is of the desired diagonal form
\begin{eqnarray*}
\hat H_{\rm BCS}=\underbrace{U_{\rm BCS}(0)}_{T=0}+\underbrace{\sum_{\vk\sigma i}
E_{\vk i}\hat\alpha_{\vk\sigma i}^\dagger\hat\alpha_{\vk\sigma i}}_{T>0}
\end{eqnarray*}
which allows for the interpretation of the operators $\hat\alpha_{\vk\sigma i}^\dagger$ and
$\hat\alpha_{\vk\sigma i}$ as ones to create and annihilate a fermionic elementary (thermal)
excitation in a quantum state $|\vk,\sigma, i\rangle$, a so--called
Bogoliubov--Valatin quasiparticle (BVQP). The quantity $U_{\rm BCS}(0)$ denotes the qround state energy, whereas
 $E_{\vk i}$ represents the energy spectrum of the BVQP on the two bands $i=1,2$.
The statistical physics of the excitation gas is exclusively describable by the BVQP
Fermi--Dirac distribution
\begin{eqnarray}
\nu_{\vk i}=\nu(E_{\vk i})=\langle\hat\alpha_{\vk\sigma i}^\dagger\hat\alpha_{\vk\sigma i}\rangle=
\frac{1}{\exp{\left(\frac{E_{\vk i}}{\kB T}\right)}+1}
\end{eqnarray}
and its derivative
\begin{eqnarray}
y_{\vk i} &\equiv& -\frac{\partial\nu_{\vk i}}{\partial E_{\vk i}}=
\frac{1}{4\kB T}\frac{1}{\cosh^2\left(\frac{E_{\vk i}}{2\kB T}\right)}
\end{eqnarray}
which is known as the so--called Yosida kernel, since it generates the band--selected Yosida functions~\cite{YOSIDA1958}
\begin{eqnarray}
Y_i(\hat\vp,T)&=&\int\limits_{-\infty}^\infty d\xi_{\vp i} y_{\vp i}\stackrel{\Delta_\vp=\Delta}{\ =\ }Y_i(T)\ .
\end{eqnarray}
In global equilibrium, the electronic distribution functions $n_{\vk i}^0$ and $g_{\vk i}$
can be evaluated by expressing the ordinary electron operators $\hat c_{\vk i}$ and
$\hat c_{\vk i}^\dagger$ through the BVQP operators $\hat \alpha_{\vk i}$ and $\hat \alpha_{\vk i}^\dagger$
using the Bogoliubov--Valatin transformation method with the result
\begin{eqnarray}
n_{\vk i}&=&\langle\hat c_{\vk\sigma i}^\dagger\hat c_{\vk\sigma i}\rangle=u_{\vk i}^2\nu_{\vk i}+v_{\vk i}^2[1-\nu_{\vk i}]
=\frac{1}{2}-\xi_{\vk i}\theta_{\vk i} \\
\label{eq:BVTg}
g_{\vk i}&=&\langle\hat c_{-\vk\downarrow i}\hat c_{\vk\uparrow i}\rangle
=-\Delta_{\vk i}\theta_{\vk i} \ \ \ ; \ \ \ \theta_{\vk i}\equiv\frac{1}{2E_{\vk i}}\tanh\frac{E_{\vk i}}{2\kB T}
\end{eqnarray}
It is convenient to combine $n_\vk$ and $g_\vk$ in an equilibrium density matrix in two--band Nambu space:
\begin{eqnarray*}
\underline{n}_{\vk i}^0&=&
\left(\begin{array}{cc}
n_{\vk i}   &      g_{\vk i} \\
g_{\vk i}^* &    1-n_{-\vk i}
\end{array} \right)
=\frac{1}{2}\left(\begin{array}{cc}
1   &  0 \\
0   &  1
\end{array} \right)-\left(\begin{array}{cc}
\xi_{\vk i}   &  \Delta_{\vk i} \\
\Delta_{\vk i}^* &   -\xi_{\vk i}
\end{array} \right)\theta_{\vk i}
=\frac{1}{2}\underline{1}-\underline{\xi}_{\vk i}\theta_{\vk i}
\end{eqnarray*}
Note that the Bogoliubov--Valatin matrix $\underline{U}_\vk$ also diagonalizes $\underline{n}_\vk$ with the result
\begin{eqnarray*}
\underline{U}_{\vk i}^\dagger\cdot\underline{n}_{\vk i}^0\cdot\underline{U}_{\vk i}
=\left( \begin{array}{cc}
\nu_{\vk i}  & 0 \\
 0 &1-\nu_{-\vk i}
\end{array} \right)
=\left( \begin{array}{cc}
\nu\left(E_{\vk i}\right)  & 0                   \\
 0              & \nu\left(-E_{\vk i}\right)
\end{array} \right)\ .
\end{eqnarray*}
\subsection{Equilibrium gap equations in two--band superconductors}
\label{sec:EquilGapEq}
The energy gaps $\Delta_{\vk i}$, are related to the Gor'kov amplitudes $g_{\vk i}$ through the set of
coupled self--consistency equations (\ref{eq:EnGap}). In strict analogy to the one--band case,
we now have to choose {\it three} weak coupling BCS model pairing interactions, which are introduced in the standard
factorizable way:
\begin{eqnarray}
\label{eq:BCSModelPairInt}
\Gamma_{\vk\vp}^{(ij)}&=&\left\{
                                        \begin{array}{lp{0.5cm}l}
                                        -\Gamma_{ij} & ; & {\rm for}\ |\xi_{\vk i}|,\ |\xi_{\vp j}|<\epsilon_0, \ i,j=1,2 \\
                                        &&\\
                                         0 & ; & {\rm otherwise}
                                         \end{array}
                                \right.
\end{eqnarray}
Here $\epsilon_0$ is a characteristic cut--off energy. Inserting this weak coupling BCS model form (\ref{eq:BCSModelPairInt}) for the pairing 
interactions $\Gamma_{\vk\vp}^{(ij)}$, one may identify
$\Delta_{\vk i}=\Delta_i$ and write
\begin{eqnarray*}
\Delta_i=-\sum_j\Gamma_{ij}{\sum_\vp}^\prime g_{\vp j}
\end{eqnarray*}
where we have defined
\begin{eqnarray*}
{\sum_\vp}^\prime A_{\vp i}=N_i(0)\int\limits_{-\epsilon_0}^{\epsilon_0}d\xi_{\vp i}A_{\vp i}
\end{eqnarray*}
with $N_i(0)$ the density of states for one spin projection on the $i$--th band.
Now using the result of the Bogoliubov--Valatin transformation from equation (\ref{eq:BVTg})
we arrive at
\begin{eqnarray*}
\Delta_i&=&\sum_j\Gamma_{ij}{\sum_\vp}^\prime \theta_{\vp j}\Delta_j \\
&=&\sum_j\underbrace{\Gamma_{ij}N_j(0)}_{\equiv\lambda_{ij}}
\underbrace{\int\limits_{-\epsilon_0}^{\epsilon_0}d\xi_{\vp j}\theta_{\vp j}}_{\equiv\Xi_j}\Delta_j\ . 
\end{eqnarray*}
It is convenient to define here dimensionless pairing interactions
\begin{eqnarray*}
\lambda_{ij}\ \equiv\ N_j(0)\Gamma_{ij} \ \ ; \ \ \vlambda&\equiv&\left( \begin{array}{cc} \lambda_{11}  & \lambda_{12} \\
\lambda_{12}  & \lambda_{22}   \end{array} \right)
\end{eqnarray*}
with $\vlambda$ the symmetric pairing interaction matrix and integrals
\begin{eqnarray}
\label{eq:SigInt}
\Xi_i&\equiv&\int\limits_{-\epsilon_0}^{\epsilon_0}d\xi_{\vp i} \theta_{\vp i}
=L(T)-P_i(T)
\end{eqnarray}
with 
\begin{eqnarray*}
  L(T)&\equiv&\ln\frac{2\epsilon_0{\rm e}^\gamma}{\pi\kB T}\ \ \ \ \ \ ;\ \ \gamma=0.5777(2)\ \text{\ (Euler's\ constant)}\\
P_i(T)&\equiv&\int\limits_{-\epsilon_0}^{\epsilon_0}d\xi_{\vp i}
\left(\frac{\tanh\frac{\xi_{\vp i}}{2\kB T}}{2\xi_{\vp i}}-\frac{\tanh\frac{E_{\vp i}}{2\kB T}}{2E_{\vp i}}\right)\ .
\end{eqnarray*}
Note, that the integrals (\ref{eq:SigInt}) take very simple forms in two different limiting cases:
\begin{eqnarray}
\label{eq:SigLimits}
\Xi_i&=&\left\{
        \begin{array}{lp{0.5cm}l l}
            \ln\frac{2\epsilon_0}{\Delta_i(0)} & ; & T\to 0 & (\text{low--}T\text{\ regime})\\
            &&\\
            L(T)-\frac{7\zeta(3)}{8}\frac{\Delta_i^2(T)}{(\pi\kB T)^2} & ; & T\to\Tc & ({\rm GL\ regime})
        \end{array}
\right.
\end{eqnarray}
Now using the definition (\ref{eq:SigInt}) we are able to write the equilibrium gap equation in a way reminiscent of the one--band case:
\begin{eqnarray}
\label{eq:EquilGapEq}
0&=&\sum_j\left\{-\delta_{ij}+\lambda_{ij}\Xi_j\right\} \Delta_j \ \ \Leftrightarrow \ \
0\ =\ \left\{-\vlambda^{-1}+\left(\begin{array}{cc}
\Xi_1  &  0 \\ 0  &  \Xi_2\end{array} \right)\right\}
\cdot\left( \begin{array}{c}
\Delta_1  \\ \Delta_2 \\
\end{array} \right)\ .
\end{eqnarray}
The coupled gap equations can be rewritten in the form:
\begin{eqnarray*}
\Xi_1&=&\frac{1}{\lambda}\left[\lambda_{22}-\lambda_{12}r(T)\right] \ \ ; \ \ r(T)\equiv\frac{\Delta_2(T)}{\Delta_1(T)} \\
\Xi_2&=&\frac{1}{\lambda}\left[\lambda_{11}-\lambda_{12}t(T)\right] \ \ ; \ \ t(T)\equiv\frac{\Delta_1(T)}{\Delta_2(T)}=\frac{1}{r(T)}
\end{eqnarray*}
Taking the difference of these coupled equations turns out to be a convenient starting point for the
calculation of the unknown ratio $r(T) = 1/t(T)$. Defining
\begin{eqnarray}
\label{eq:LeggettParam}
\begin{aligned}
a&\ \equiv\ \frac{\lambda_{22}-\lambda_{11}}{\lambda} \\
\gamma_{\rm L}&\ \equiv\ \frac{\lambda_{12}}{\lambda} \\
\lambda&\ \equiv\ {\rm det}\vlambda=\lambda_{11}\lambda_{22}-\lambda_{12}^2
\end{aligned}
\end{eqnarray}
with the quantity $\gamma_{\rm L}$ being referred to as the {\it Leggett coupling}~\cite{LEGGETT1966}, we obtain
\begin{eqnarray*}
\Xi_1-\Xi_2&=&a-\gamma_{\rm L}\left[r(T)-t(T)\right]\ .
\end{eqnarray*}
In what follows we assume without restricting generality that $\Delta_2(T)>\Delta_1(T)$, or alternatively $r(T)~>~1$.
\subsubsection{Determination of the transition temperature}
At the transition temperature $\Tc$ we may state that
\begin{eqnarray*}
\Xi_1(\Tc)=\Xi_2(\Tc)=L(\Tc)=\frac{1}{\lambda}\left[\lambda_{22}-\lambda_{12}r(\Tc)\right]=\frac{1}{\lambda}\left[\lambda_{11}-\lambda_{12}t(\Tc)\right]\ .
\end{eqnarray*}
This condition yields quadratic equations in $r(\Tc)$ and $t(\Tc)$,
\begin{eqnarray*}
r^2(\Tc)-\mu r(\Tc)-1&=&0 \\
t^2(\Tc)+\mu t(\Tc)-1&=&0
\end{eqnarray*}
which have the solutions
\begin{eqnarray}
 \label{eq:rtGL}
\begin{aligned}
r(\Tc)&\ =\ \frac{\mu}{2}+\sqrt{\frac{\mu^2}{4}+1} \ \ \ \ \ \ ; \ \ \mu=\frac{a}{\gamma_{\rm L}} \\
t(\Tc)&\ =\ -\frac{\mu}{2}+\sqrt{\frac{\mu^2}{4}+1}
\end{aligned}
\end{eqnarray}
and the obvious condition $r(\Tc)\cdot t(\Tc)=1$ holds. From this, the transition temperature $\Tc$ is obtained in the
following form
\begin{eqnarray}
\label{eq:TrTemp}
\pi\kB\Tc=2\epsilon_0{\rm e}^\gamma  {\rm e}^{-\frac{1}{2\lambda}[\lambda_{22}+\lambda_{11}-
\sqrt{(\lambda_{22}-\lambda_{11})^2+4\lambda_{12}^2}]}\ ,
\end{eqnarray}
which coincides with the result of {\it Suhl}, {\it Matthias} and {\it Walker}~\cite{SMANDW1959} derived as early as 1959. 
Note, that in case of two decoupled gaps ($\lambda_{12}=0$) one is left with two different BCS transition temperatures:
\begin{eqnarray*}
 \pi\kB T_{{\rm c}(i)}=2\epsilon_0{\rm e}^\gamma  {\rm e}^{-\frac{1}{\lambda_{ii}}}\ ,\ i=1,2
\end{eqnarray*}
A careful analysis of Eq. (\ref{eq:TrTemp}) shows, 
that the transition temperature $\Tc(\lambda_{12})$ is always larger than the largest of the values $\Tc(\lambda_{12}=0)$. 
This fact has actually been discussed in previous publications~\cite{KONDO1963, RAINER1968}. 
In order to visualize this statement, we have plotted the quantity $T_{{\rm c}}(\lambda_{12})/T_{{\rm c}2}$ vs. $\lambda_{12}$ 
in Fig.~\ref{fig:transtemp}. For calculations we used the model with the intra--band coupling 
constants $\lambda_{11}=0.12$ and $\lambda_{22}=0.19$, which is suitable for MgB$_2$--like two--band superconductors.
The figure clearly shows the monotonic increase of $\Tc(\lambda_{12})$ away from its value $T_{{\rm c}2}$ at $\lambda_{12}=0$.
\begin{figure}[!htb]
\centering
	\begin{overpic}[width=0.61\textwidth]{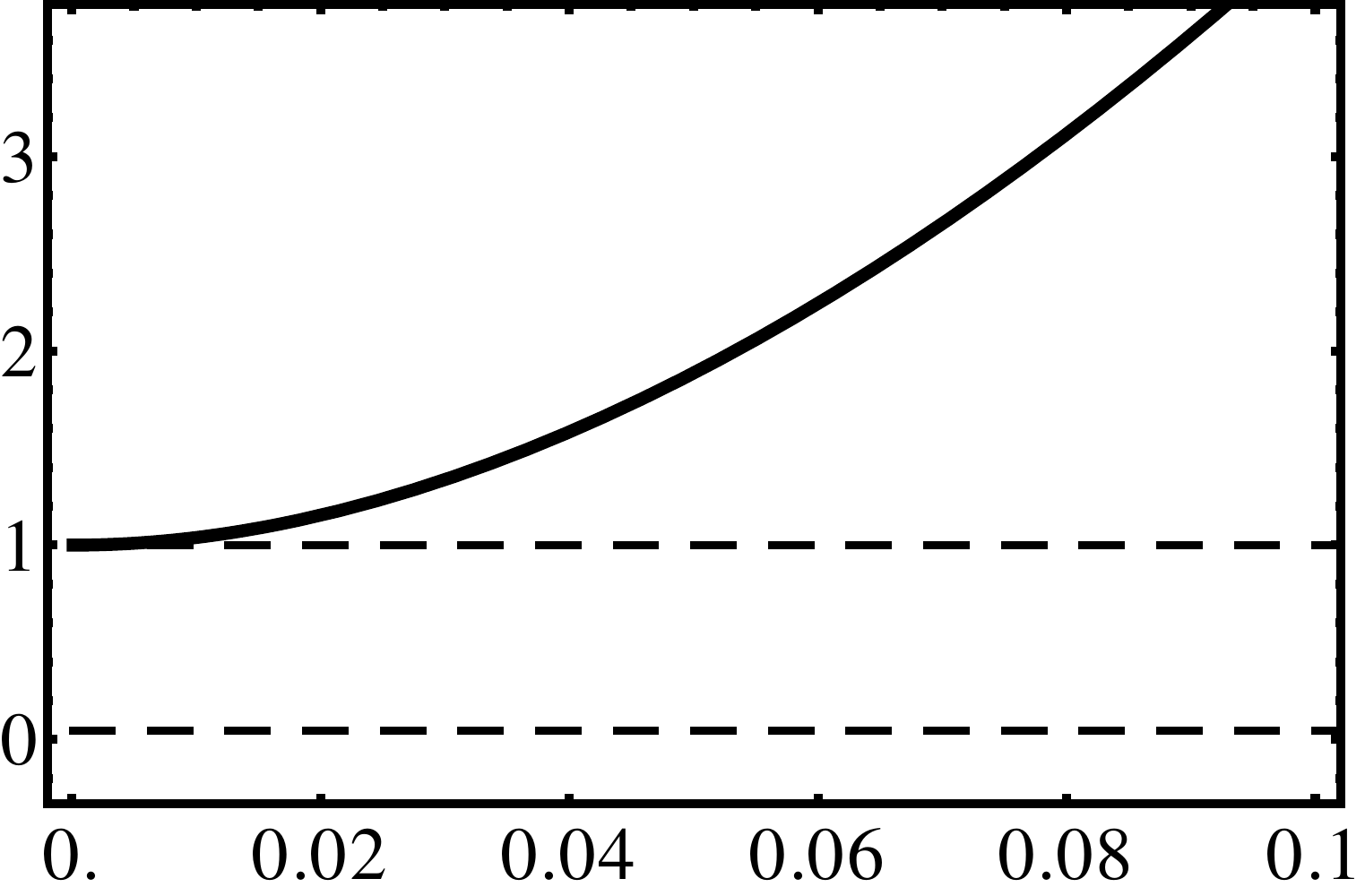} 
	\put(50,-7){\Large $\lambda_{12}$}
	\put(-5,18){\Large \begin{rotate}{90} $T_{{\rm c}(i)}/T_{{\rm c}2},\ i=1,2$\end{rotate}}
	\put(59,48){\Large $\Tc$}
	\put(59,27.5){\Large $T_{{\rm c}2}$}
	\put(59,14){\Large $T_{{\rm c}1}$}	
	\end{overpic}
\captionsetup{width=.85\textwidth,font=small,labelfont=bf, skip=30pt}
\caption{Normalized transition temperature as a function of inter--band coupling constant $\lambda_{12}$.
The solid line represents the transition temperature $\Tc$ of a two--band superconductor with intra--band
coupling constants $\lambda_{11}=0.12$ and $\lambda_{22}=0.19$. The dashed lines show the case of decoupled bands
with two distinct transition temperatures $T_{{\rm c}1}$ and $T_{{\rm c}2}$ for $\lambda_{11}=0.12$ and 
$\lambda_{22}=0.19$ respectively.}
\label{fig:transtemp}
\end{figure}
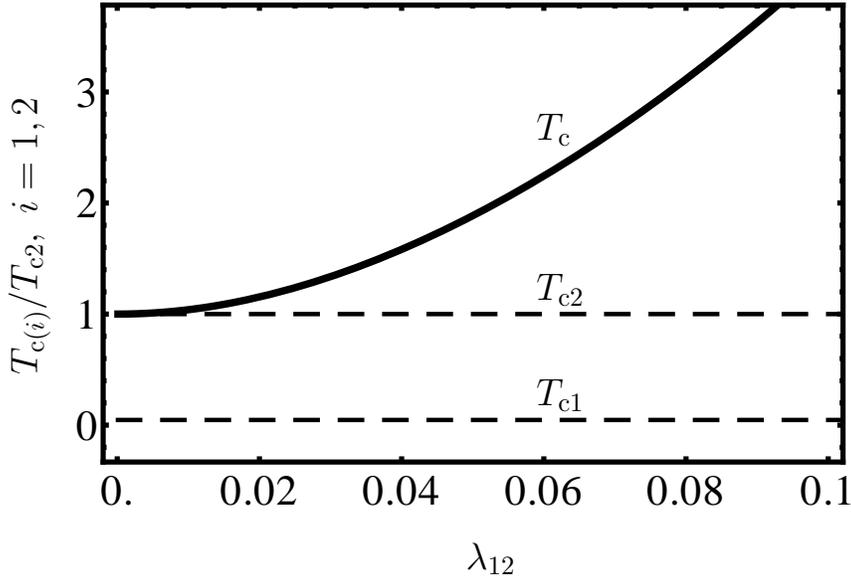
\subsubsection{Determination of the gaps in the Ginzburg--Landau regime}
Inserting the GL expansions (\ref{eq:SigLimits}) into the eqs.(\ref{eq:EquilGapEq}) for $\Xi_i(T)$  and solving for $\Delta_i(T)$, we arrive at
\begin{eqnarray}
\label{eq:EquilGapEq_GL}
\begin{aligned}
\frac{\Delta_1^2(T)}{(\pi\kB T)^2}&\ =\ \frac{8}{7\zeta(3)}\left\{\ln\frac{\Tc}{T}+\gamma_{\rm L}\left[r(T)-r(\Tc)\right]\right\} \\
\frac{\Delta_2^2(T)}{(\pi\kB T)^2}&\ =\ \frac{8}{7\zeta(3)}\left\{\ln\frac{\Tc}{T}+\gamma_{\rm L}\left[t(T)-t(\Tc)\right]\right\}\ .
\end{aligned}
\end{eqnarray}
Multiplication of the first of these equations with $r^2(T)$ yields
\begin{eqnarray*}
\ln\frac{\Tc}{T}+\gamma_{\rm L}\left[t(T)-t(\Tc)\right]=r^2(T)\left\{
\ln\frac{\Tc}{T}+\gamma_{\rm L}\left[r(T)-r(\Tc)\right]\right\}
\end{eqnarray*}
This can easily be rearranged to the form of a pair of fourth order equations for $r(T)$
and $t(T)$, respectively:
\begin{eqnarray*}
r^4(T)+\alpha r^3(T)-\beta r(T)-1&=&0 \\
t^4(T)+\beta t^3(T)-\alpha t(T)-1&=&0 \\
\end{eqnarray*}
where we have introduced
\begin{eqnarray*}
\alpha&\equiv&\frac{1}{\gamma_{\rm L}}\left[\ln\frac{\Tc}{T}-\gamma_{\rm L}r(\Tc)\right] \\
\beta&\equiv&\frac{1}{\gamma_{\rm L}}\left[\ln\frac{\Tc}{T}-\frac{\gamma_{\rm L}}{r(\Tc)}\right]  \\
\end{eqnarray*}
In order to solve these equations, we used Newton's method.
We started from the known ratios $r_0(T)=r(\Tc)$, $t_0(T)=t(\Tc)$ and obtained by means of this procedure
the first order approximation $r_1(T)$, $t_1(T)$ in the following obvious way
\begin{eqnarray}
\label{eq:GapEqSolr_1}
\begin{aligned}
r_{1}(T)&=r(\Tc)-\frac{r^4(\Tc)+\alpha r^3(\Tc)-\beta r(\Tc)-1}{4r^3(\Tc)+3\alpha r^2(\Tc)-\beta} \\
&=r(\Tc)-\frac{r^2(\Tc)[r^2(\Tc)-1]\ln\frac{\Tc}{T}}{\gamma_{\rm L}[1+r^4(\Tc)]+r(\Tc)[3r^2(\Tc)-1]\ln\frac{\Tc}{T}} \\
t_{1}(T)&=t(\Tc)-\frac{t^4(\Tc)+\beta t^3(\Tc)-\alpha t(\Tc)-1}{4t^3(\Tc)+3\beta t^2(\Tc)-\alpha} \\
&=t(\Tc)-\frac{t^2(\Tc)[t^2(\Tc)-1]\ln\frac{\Tc}{T}}{\gamma_{\rm L}[1+t^4(\Tc)]+t(\Tc)[3t^2(\Tc)-1]\ln\frac{\Tc}{T}}
\end{aligned}
\end{eqnarray}
Note that the temperature dependence of the gap ratios $r(T)$, $t(T)$ differs from $r(\Tc)$, $t(\Tc)$ only in the case
$r(\Tc)\not=1$, $t(\Tc)\not=1$. \\
\vspace{0.1mm}\\
The result (\ref{eq:GapEqSolr_1}) can be iteratively improved as follows:
\begin{eqnarray*}
r_{n+1}(T)&=&r_n(T)-\frac{r_n^4(T)+\alpha r_n^3(T)-\beta r_n(T)-1}{4r_n^3(T)+3\alpha r_n^2(T)-\beta} \ \ ; \ \ n\geq 1 \\
t_{n+1}(T)&=&t_n(T)-\frac{t_n^4(T)+\beta t_n^3(T)-\alpha t_n(T)-1}{4t_n^3(T)+3\beta t_n^2(T)-\alpha} \ \ ; \ \ n\geq 1
\end{eqnarray*}
Eventually, one arrives at the exact result for the ratio $r(T)$ and $t(T)$ in the limiting form
\begin{eqnarray*}
r(T)&=&\lim_{n\to\infty}r_n(T) \\
t(T)&=&\lim_{n\to\infty}t_n(T)
\end{eqnarray*}
Now the known results for $r(T)$ and $t(T)$ can be inserted back into (\ref{eq:EquilGapEq_GL}) to yield the final results for the two
gaps $\Delta_1(T)$ and $\Delta_2(T)$. Using the first order approximations (\ref{eq:GapEqSolr_1}), i.e. 
$r(T)\approx r_1(T)$ and $t(T)\approx t_1(T)$, we obtain for the temperature dependence of the gaps in the Ginzburg--Landau regime
\begin{eqnarray}
\begin{aligned}
\frac{\Delta_1^2(T)}{(\pi\kB T)^2}&\ =\ \frac{\ln\frac{\Tc}{T}+\frac{\gamma_{\rm L}[1+r^2(\Tc)]}{r(\Tc)[3r^2(\Tc)-1]}}
{\ln\frac{\Tc}{T}+\frac{\gamma_{\rm L}[1+r^4(\Tc)]}{r(\Tc)[3r^2(\Tc)-1]}}\ \frac{8}{7\zeta(3)}\ln\frac{\Tc}{T} \\
\frac{\Delta_2^2(T)}{(\pi\kB T)^2}&\ =\ \frac{\ln\frac{\Tc}{T}+\frac{\gamma_{\rm L}[1+t^2(\Tc)]}{t(\Tc)[3t^2(\Tc)-1]}}
{\ln\frac{\Tc}{T}+\frac{\gamma_{\rm L}[1+t^4(\Tc)]}{t(\Tc)[3t^2(\Tc)-1]}}\ \frac{8}{7\zeta(3)}\ln\frac{\Tc}{T}
\end{aligned}
\end{eqnarray}
As expected in the limit of vanishing Leggett coupling
$\gamma_{\rm L}\to 0$, the two gaps $\Delta_1(T)$ and $\Delta_2(T)$
show BCS temperature dependence in the Ginzburg--Landau regime
\begin{eqnarray*}
 \lim_{\gamma_{\rm L}\to 0}\frac{\Delta_i^2(T)}{(\pi\kB T)^2}
&=&\frac{8}{7\zeta(3)}\ln\frac{T_{{\rm c}i}}{T} \ \ ; \ \ i=1,2\ .
\end{eqnarray*}
\subsubsection{Determination of the gaps at zero temperature}
At low temperature, we may use equation (\ref{eq:SigLimits}) for $\Xi_i(T)$ and immediately obtain from (\ref{eq:EquilGapEq})
a pair of transcendental equations for the ratios $r(0)$ and $t(0)$:
\begin{eqnarray}
\label{eq:TransEqrAndt}
\begin{aligned}
\ln r(0)&\ =\ a-\gamma_{\rm L}\left[r(0)-\frac{1}{r(0)}\right] \\
\ln t(0)&\ =\ -a-\gamma_{\rm L}\left[t(0)-\frac{1}{t(0)}\right]\ .
\end{aligned}
\end{eqnarray}
The two gaps $\Delta_i(0)$ at $T=0$ can then be expressed through $r(0)$ and $t(0)$ in the form
\begin{eqnarray}
\label{eq:EquilGapEq_LowT}
\begin{aligned}
\frac{\Delta_1(0)}{\kB\Tc}&=\frac{\pi}{{\rm e}^\gamma}{\rm e}^{-\gamma_{\rm L}\left[r(\Tc)-r(0)\right]}\\
\frac{\Delta_2(0)}{\kB\Tc}&=\frac{\pi}{{\rm e}^\gamma}{\rm e}^{-\gamma_{\rm L}\left[t(\Tc)-t(0)\right]}\ .
\end{aligned}
\end{eqnarray}
Clearly, the modification of the BCS result for the zero temperature gap consists in enlargement factor involving
$r(0)$ and a reduction factor involving $t(0)$ for the two gaps $\Delta_1(0)$ and $\Delta_2(0)$, respectively.
As a consequence, the zero--temperature gaps (\ref{eq:EquilGapEq_LowT}) of two--band (two--gap) superconductors are seen to
represent the first non--universal BCS--M\"uhlschlegel parameter (the second being the specific heat discontinuity, to be discussed
in section \ref{sec:SpHeatDiscontinuity}). It remains to calculate $r(0)$. The transcendental equation (\ref{eq:TransEqrAndt}) for the determination of
the ratio $r(0)$
can be solved, if $r(0)$ is close to one. Then one may expand the logarithm $\ln r(0)\approx -[1-r(0)]$ and obtains the approximate result
\begin{eqnarray*}
r_1(0)&=&\frac{1}{2(1+\gamma_{\rm L})}\left\{1+a+\sqrt{(1+a)^2+4\gamma_{\rm L}(1+\gamma_{\rm L})}\right\} \\
&\stackrel{a\ll 1}{=}&1+\frac{a}{1+2\gamma_{\rm L}}+\dots
\end{eqnarray*}
This result can be iteratively improved using Newton's procedure
\begin{eqnarray*}
r_{n+1}(0)=r_n(0)-\frac{\ln r_n(0)+\gamma_{\rm L}\left[r_n(0)-\frac{1}{r_n(0)}\right]-a}
{\frac{1}{r_n(0)}+\gamma_{\rm L}\left[1+\frac{1}{r^2_n(0)}\right]}  \ \ ; \ \ n\geq 1
\end{eqnarray*}
and we may state that we have found an exact result for $r(0)$:
\begin{eqnarray}
\label{eq:rtLT}
r(0)=\lim_{n\to\infty}r_n(0)
\end{eqnarray}
%
%
\subsubsection{Numerical results for the gaps}
This section is devoted to a discussion of the numerical solutions of the coupled gap equations (\ref{eq:EquilGapEq})
at arbitrary temperatures, as compared with available analytical solutions in both the Ginzburg--Landau and the zero temperature limit.
The intra--band pair coupling constants were again chosen to be $\lambda_{11}=0.12$ and $\lambda_{22}=0.19$. 
In Figs.~\ref{fig:gaps_1912005} and~\ref{fig:gaps_191207}, we have plotted the normalized gap functions $\Delta_{i}^{2}(T)/\Delta_2^2(0)$, 
 $i=1$ (\textit{right axis}) and $2$ (\textit{left axis}) as a function of reduced temperature $T/\Tc$ 
for values of $\lambda_{12}=0.005$ (\ref{fig:gaps_1912005}) and $\lambda_{12}=0.07$ (\ref{fig:gaps_191207}). 
As can be immediately seen from these figures, the large gap $\Delta_2^2(T)$ shows a 
BCS--like behavior when plotted against the reduced temperature, whereas the small gap $\Delta_1^2(T)$ 
displays a non--monotonic temperature dependence, when $\lambda_{12}$ is very small. The latter behavior can be explained
by the fact, that the gaps are nearly independent in this case.
This is reflected also in the low--$T$ 
behavior of the local response functions, which will be discussed in section~\ref{sec:LocRespFunc}. 
Also shown in Figs.~\ref{fig:gaps_1912005} and~\ref{fig:gaps_191207} 
as dashed lines are the analytical results obtained in the Ginzburg--Landau and the low--$T$ limit, respectively. 
Fig.~\ref{fig:gaps_1912XX} shows a comparison of $\Delta_i^2(T)/\Delta_2^2(0)$, $i=1,2$ for the case $\lambda_{11}=0.12$ and $\lambda_{22}=0.19$ 
and various interband coupling parameters $\lambda_{12}$, as indicated in the figure caption.
Since the temperature dependence of the large gap remains nearly unaffected by the variation of the interband coupling, we 
present in Fig.~\ref{fig:gaps_1912XX} only the case of $\lambda_{12}=0.07$.
The small gap shows, however, a transition to the nearly independent BCS behavior for a sufficiently small value of $\lambda_{12}$.
\\
\begin{figure}[!htb]\ContinuedFloat*
\centering
	\begin{overpic}[width=0.65\textwidth]{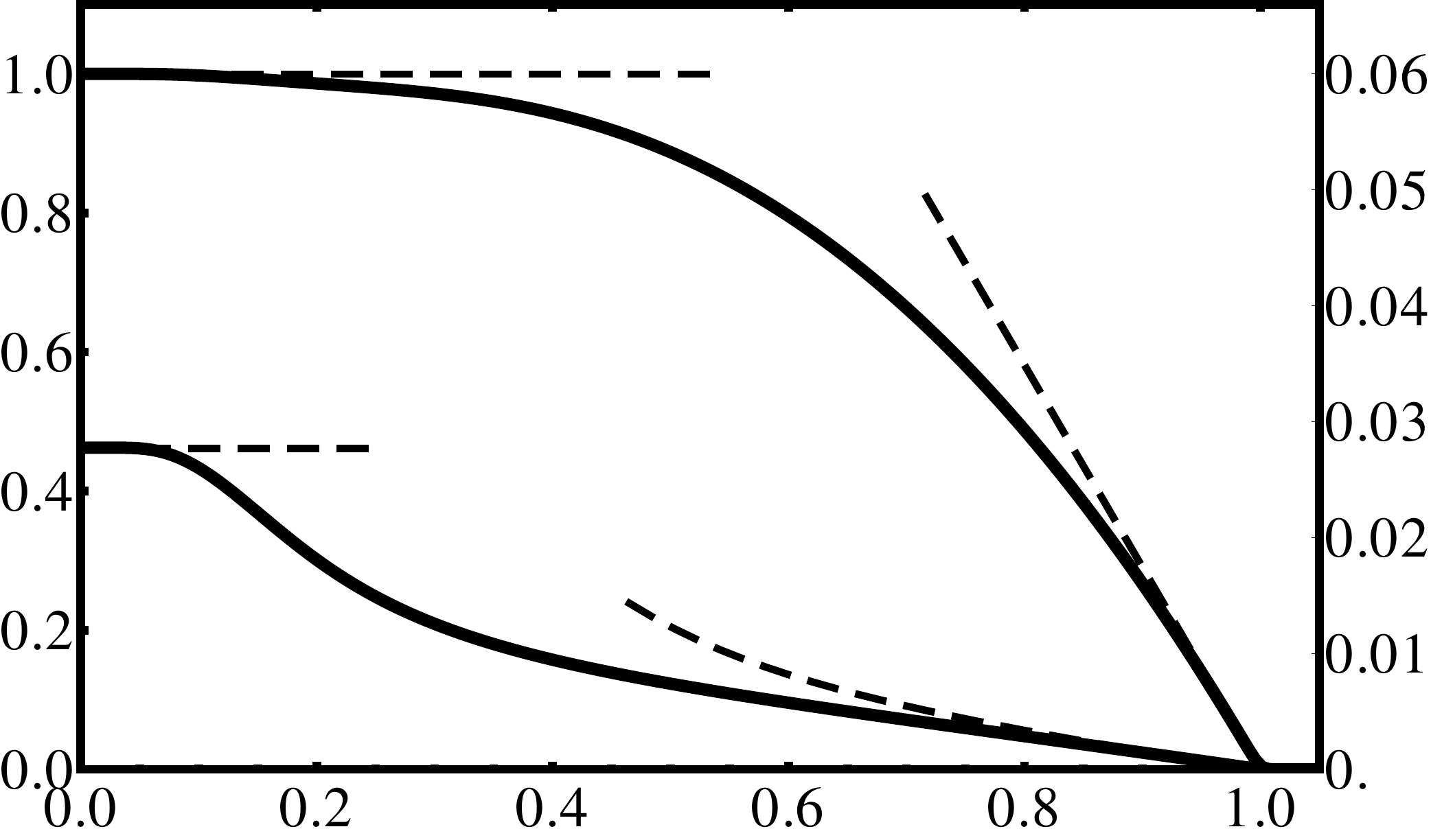} 
	\put(50,-7){\Large $T/\Tc$}
	\put(-5,22){\Large \begin{rotate}{90} $\Delta^{2}_{2}(T)/\Delta^{2}_{2}(0)$\end{rotate}}
	\put(107,22){\Large \begin{rotate}{90} $\Delta^{2}_{1}(T)/\Delta^{2}_{2}(0)$\end{rotate}}
	\put(47,40){\line(1,2){3}}
	\put(39,35){\large $\frac{\Delta^{2}_{2}(T)}{\Delta^{2}_{2}(0)}$}
	\put(16,16){\line(1,2){3}}
	\put(8,11){\large $\frac{\Delta^{2}_{1}(T)}{\Delta^{2}_{2}(0)}$}
	\end{overpic}
\captionsetup{width=.85\textwidth,font=small,labelfont=bf, skip=30pt}
\caption{Two--axis plot for the evolution of the normalized superconducting gaps $\Delta^{2}_1(T)/\Delta^{2}_2(0)$ (\textit{small gap}, right axis)
and $\Delta^{2}_2(T)/\Delta^{2}_2(0)$ (\textit{large gap}, left axis) with reduced temperature $T/\Tc$ for intra--band
coupling constants $\lambda_{11}=0.12$ and $\lambda_{22}=0.19$ and inter--band coupling constant $\lambda_{12}=0.005$.
Analytical solutions in low temperature limit and in the Ginzburg--Landau limit are illustrated by dashed lines.}
\label{fig:gaps_1912005}
\end{figure}
%
%
\begin{figure}[!htb]\ContinuedFloat
\centering
	\begin{overpic}[width=0.63\textwidth]{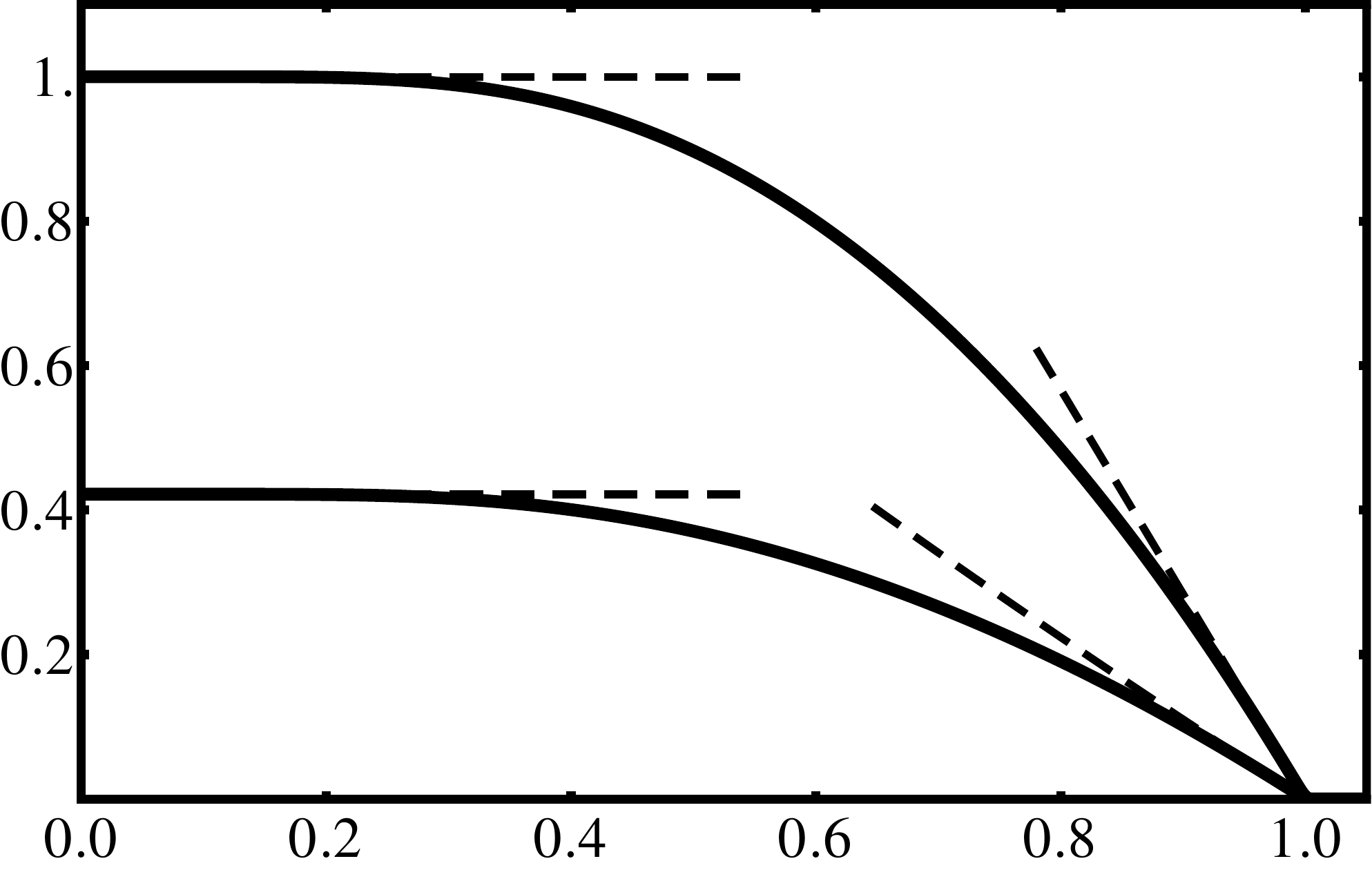} 
	\put(50,-7){\Large $T/\Tc$}
	\put(-5,12){\Large \begin{rotate}{90} $\Delta^{2}_{i}(T)/\Delta^{2}_{2}(0),\ i=1,2$\end{rotate}}
	\put(50,45){\line(1,2){3}}
	\put(42,40){\large $\frac{\Delta^{2}_{2}(T)}{\Delta^{2}_{2}(0)}$}
	\put(30,21){\line(1,2){3}}
	\put(22,16){\large $\frac{\Delta^{2}_{1}(T)}{\Delta^{2}_{2}(0)}$}
	\end{overpic}
\captionsetup{width=.85\textwidth,font=small,labelfont=bf, skip=30pt}
\caption{Evolution of the normalized superconducting gaps $\Delta^{2}_1(T)/\Delta^{2}_2(0)$ (\textit{small gap}) and $\Delta^{2}_2(T)/\Delta^{2}_2(0)$
(\textit{large gap}) with reduced temperature $T/\Tc$ for intra--band coupling constants $\lambda_{11}=0.12$ and $\lambda_{22}=0.19$ and
inter--band coupling constant $\lambda_{12}=0.07$. Analytical solutions in low temperature limit and in the Ginzburg--Landau limit are
shown as dashed lines.}
\label{fig:gaps_191207}
\end{figure}\\
\vspace{5.0cm}
\begin{figure}[!htb]\ContinuedFloat
\centering
	\begin{overpic}[width=0.633\textwidth]{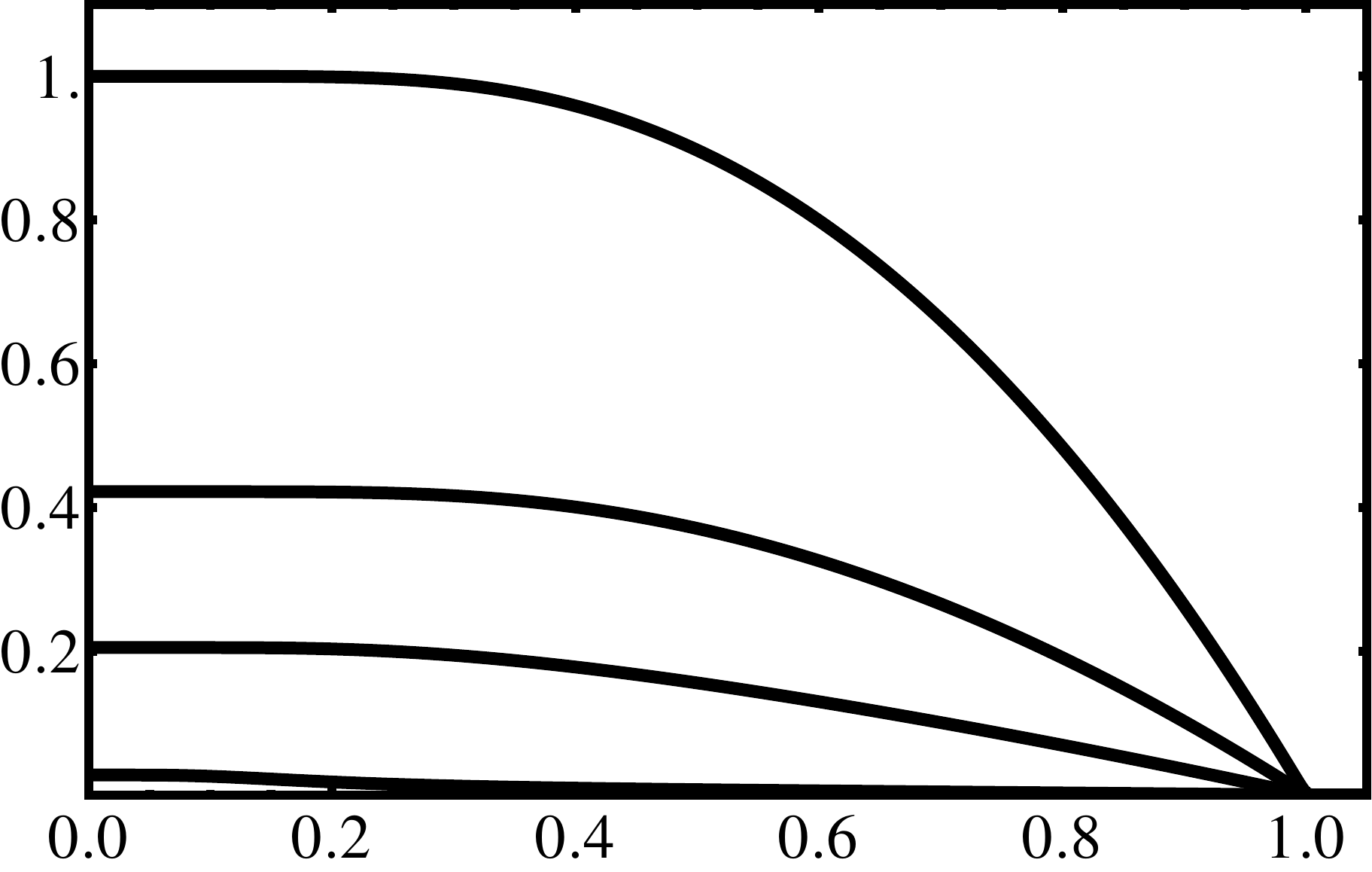} 
	\put(50,-7){\Large $T/\Tc$}
	\put(-5,12){\Large \begin{rotate}{90} $\Delta^{2}_{i}(T)/\Delta^{2}_{2}(0),\ i=1,2$\end{rotate}}
	\put(75,45){\line(-3,-2){7}}
	\put(75,45){\large $\frac{\Delta^{2}_{2}(T)}{\Delta^{2}_{2}(0)}$}
	\put(45,33){\line(-5,-2){14}}
	\put(45,33){\line(-5,-4){20}}
	\put(45,33){\line(-4,-5){21}}
	\put(10,59){(a)}
	\put(10,29){(b)}
	\put(10,18){(c)}
	\put(10,8.5){(d)}
	\put(45,33){\large $\frac{\Delta^{2}_{1}(T)}{\Delta^{2}_{2}(0)}$}
	\end{overpic}
\captionsetup{width=.85\textwidth,font=small,labelfont=bf, skip=30pt}
\caption{Evolution of the normalized superconducting gaps $\Delta^{2}_1(T)/\Delta^{2}_2(0)$ (\textit{small gap}) 
and $\Delta^{2}_2(T)/\Delta^{2}_2(0)$ (\textit{large gap}) with reduced temperature $T/\Tc$ 
for intra--band coupling constants $\lambda_{11}=0.12$ and $\lambda_{22}=0.19$ and
 various inter--band coupling constants:\\ 
 \hspace*{0.4cm} (a) $\lambda_{12}=0.07$ (for \textit{large gap}), \\
 \hspace*{0.4cm} (b) $\lambda_{12}=0.07$, (c) $\lambda_{12}=0.03$, (d) $\lambda_{12}=0.005$ (for \textit{small gap}).
 }
\label{fig:gaps_1912XX}
\end{figure}
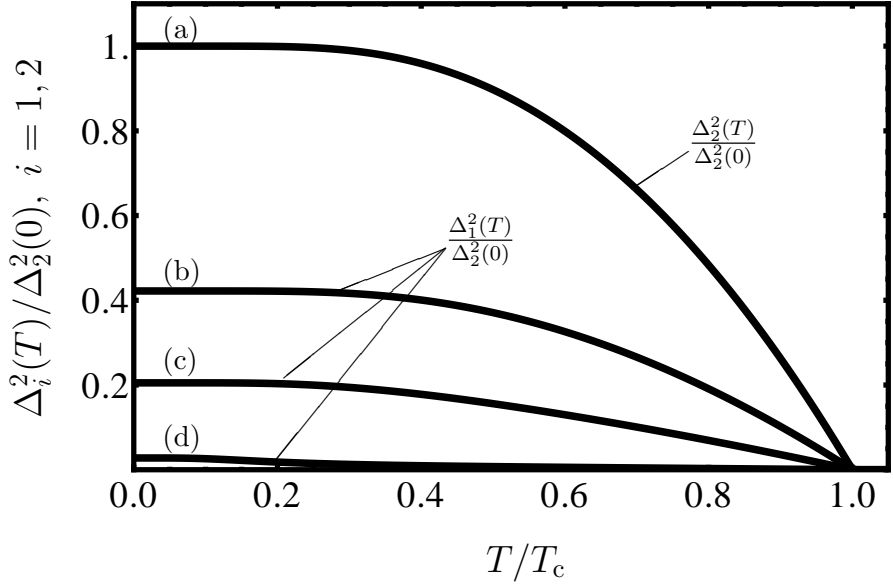
\newpage
\section{Two--band superconductors in local thermodynamic equilibrium}
\label{sec:LocRespFunc}
In what follows, we consider deviations from global thermodynamic equilibrium.
To keep things on the simplest nontrivial level, we consider perturbation potentials
$\delta\xi_{\vk i\sigma\sigma^\prime}^{\rm ext}$, which are slowly varying in time and space.
Such a situation, which corresponds to the long wavelength ($\vq\to 0$) low frequency or stationary ($\omega\to 0$)
limit, is referred to as {\it local equilibrium}, and can be described by the perturbation Hamiltonian
\begin{eqnarray}
\hat{H}^{\rm ext}&=&\frac{1}{V}\sum_{\vp\sigma\sigma' i}
\hat{c}_{\vp\sigma i}^\dagger\delta\xi_{\vp i\sigma\sigma'}^{\rm ext}\hat{c}_{\vp\sigma' i}
\end{eqnarray}
For the following discussion, we shall show that the perturbation potentials $\delta\xi_{\vp i\sigma\sigma'}^{\rm ext}$
can be written in the general form
\begin{eqnarray}
\label{eq:PerPotGenForm}
\delta\xi_{\vp i\sigma\sigma'}^{\rm ext}=\left[\vp_i\cdot\vv_i^{\rm s}
-\left(\frac{E_{\vk i}}{T}-\frac{\partial E_{\vk i}}{\partial T}\right)\delta T\right]\delta_{\sigma\sigma'}
-\frac{\gamma\hbar}{2}\vtau_{\sigma\sigma'}\cdot\vH^{\rm ext}
\end{eqnarray}
The three terms above describe the coupling of the quasiparticle system to
(i) the condensate velocity $\vv_i^{\rm s}$, to be derived below, (ii) to a local temperature change
$\delta T$ and (iii) to an external magnetic field (Zeeman coupling). In the latter term $\gamma$ denotes the gyromagnetic
ratio of electrons.
\subsection{Normal and superfluid density}
In the presence of a vector potential there will be an Amp\`ere coupling induced quasiparticle energy shift
\begin{eqnarray*}
\delta\xi_{\vk i\sigma\sigma'}^{\rm ext}
=\left\{\frac{1}{2m}\left(\vp_i-\frac{e}{c}\vA\right)^2-\frac{\vp_i^2}{2m}\right\}\delta_{\sigma\sigma'}
=-\vv_{\vk i}\cdot\frac{e}{c}\vA\delta_{\sigma\sigma'}
\end{eqnarray*}
In order to arrive at a gauge--invariant form of the Amp\`ere coupling, we perform
a gauge transformation to the vector potential of the form
\begin{eqnarray*}
-\vv_{\vk i}\cdot{\frac{e}{c}\vA}&\to&
-\vv_{\vk i}\cdot\left\{{\frac{e}{c}\vA-\frac{\hbar}{2}\vna\chi_i}\right\}\equiv\vp_i\cdot\vv^{\rm s}_i \\
\vv^{\rm s}_i&=&\frac{1}{m}\left(\frac{\hbar}{2}\vna\chi_i-\frac{e}{c}\vA\right)
\end{eqnarray*}
where the scalar functions $\chi_i$ denote the phases of the superconducting order parameters on the bands $i=1,2$,
which therefore acquire the physical meaning of {\it velocity potentials} for the superflow described by $\vv^{\rm s}_i$.
Hence we may rewrite $\delta\xi_{\vk i\sigma\sigma^\prime}^{\rm ext}$ in the convenient form of {\it Doppler shifts}:
\begin{eqnarray}
\delta\xi_{\vk i\sigma\sigma^\prime}^{\rm ext}=\vp_i\cdot\vv^{\rm s}_i\delta_{\sigma\sigma^\prime}
\end{eqnarray}
Now the supercurrent density can be written in the standard gauge--invariant quantum--mechanical form,
which can be Taylor--expanded w.r.t. the small superflow velocities $\vv^{\rm s}_i$:
\begin{eqnarray*}
\vj^{\rm s}&=&\sum_{i=1}^2\left[n_i\vv^{\rm s}_i+\frac{1}{V}\sum_{\vp\sigma}\vv_{\vp i}\nu\left(E_{\vp i}+\vp_i\cdot\vv^{\rm s}_i\right)\right]\\
&=&\sum_{i=1}^2\left[n_i\vv^{\rm s}_i+\frac{1}{V}\sum_{\vp\sigma}\vv_{\vp i}\left\{\nu\left(E_{\vp i}\right)
+\frac{\partial\nu\left(E_{\vp i}\right)}{\partial E_{\vp i}}\vp_i\cdot\vv^{\rm s}_i\right\}\right] \\
&=&\sum_{i=1}^2\left[n_i\vv^{\rm s}_i-\frac{1}{V}\sum_{\vp\sigma}y_{\vp i}\vv_{\vp i}\left(\vp_i\cdot\vv^{\rm s}_i\right)\right]
\end{eqnarray*}
This immediately implies the definition of the band--selected {\it normal fluid density} tensor in the form
\begin{eqnarray}
\label{eq:NormFlDens}
n^{\rm n}_{\mu\nu i}&=&\sum_{\vp\sigma}\frac{p_{\mu i}p_{\nu i}}{m}y_{\vp i}
=N_{{\rm F}i}\lela\overbrace{\int\limits_{-\mu_i}^\infty d\xi_{\vp i}y_{\vp i}}^{=Y_i(\hat\vp,T)}\frac{p_{\mu i} p_{\nu i}}{m}\rira_{{\rm FS}i} \\
&=&\underbrace{N_{{\rm F}i}\frac{p_{{\rm F}i}^2}{m}}_{=3n_i}\lela Y_i(\hat\vp,T)\hat\vp_{\mu} \hat\vp_{\nu}\rira_{{\rm FS}i}
=n_i\ 3\lela Y_i(\hat\vp,T)\hat\vp_{\mu} \hat\vp_{\nu}\rira_{{\rm FS}i}
\stackrel{\Delta_\vp=\Delta}{=}n_iY_i(T)\delta_{\mu\nu}
\end{eqnarray}
Note that $\lela\dots\rira_{{\rm FS}i}$ denotes the average over the $i$--th Fermi surface and $n=n_1+n_2$.
Correspondingly, the band--selected {\it superfluid density tensor} can be defined as
\begin{eqnarray*}
n^{\rm s}_{\mu\nu i}&=&n_i\delta_{\mu\nu}-n^{\rm n}_{\mu\nu i}
=n_i\left[\delta_{\mu\nu}-3\lela Y_i(\hat\vp)\hat\vp_{\mu} \hat\vp_{\nu}\rira_{{\rm FS}i}\right]
\stackrel{\Delta_\vp=\Delta}{=}n_i\left[1-Y_i(T)\right]\delta_{\mu\nu}
\end{eqnarray*}
Finally, we may summarize our result for the supercurrent response in two--band superconductors as follows:
\begin{eqnarray}
j^{\rm s}_\mu=\sum_{i=1}^2\left(n_i\delta_{\mu\nu}-n^{\rm n}_{\mu\nu i}\right)v^{\rm s}_{\nu i}
=\sum_{i=1}^2n^{\rm s}_{\mu\nu i}v^{\rm s}_{\nu i}
\stackrel{\Delta_\vp=\Delta}{=}\sum_{i=1}^2n_i^{\rm s}v^{\rm s}_{\mu i}
\end{eqnarray}
\subsubsection{The London--BCS magnetic penetration depth}
In order to derive an expression for the London--BCS magnetic penetration depth $\lambda_{\rm L}(T)$ we start from
(\ref{eq:PerPotGenForm}) and write
\begin{eqnarray*}
\vj_e^{\rm s}\equiv e\vj^{\rm s}=en^{\rm s}\vv^{\rm s} \ \ ; \ \ n^{\rm s}\equiv\sum_{i=1}^2n^{\rm s}_i
\end{eqnarray*}
As a next step we use Amp\`ere's law $\vna\times\vB=4\pi\vj_e^{\rm s}/c$ to derive the screening differential equation
$\vna^2\vB=\vB/\lambda_{\rm L}^2$, with
\begin{eqnarray}
\label{eq:MagnPenDepth}
\lambda_{\rm L}^2(T)=\frac{mc^2}{4\pi n^{\rm s}(T)e^2}
\end{eqnarray}
representing the London--BCS magnetic penetration depth.
In Fig.~\ref{fig:pen_depth} we have plotted the dependence of the normalized magnetic penetration depth $\lambda_{\rm L}(T)/\lambda_{\rm L}(0)$ 
on reduced temperature $T/\Tc$ for the set of intra--band pair coupling constants $\lambda_{11}=0.12$ and $\lambda_{22}=0.19$ 
and two values for $\lambda_{12}=0.005\ (i)$ and $\lambda_{12}=0.07\ (ii)$. 
For the lower value of $\lambda_{12}$, the magnetic penetration depth shows non--monotonic behavior, 
which can be traced back to thermal activation processes associated with the smaller gap $\Delta_1(T)$.
%
%
%
\begin{figure}[!htb]
\centering
	\begin{overpic}[width=0.6\textwidth]{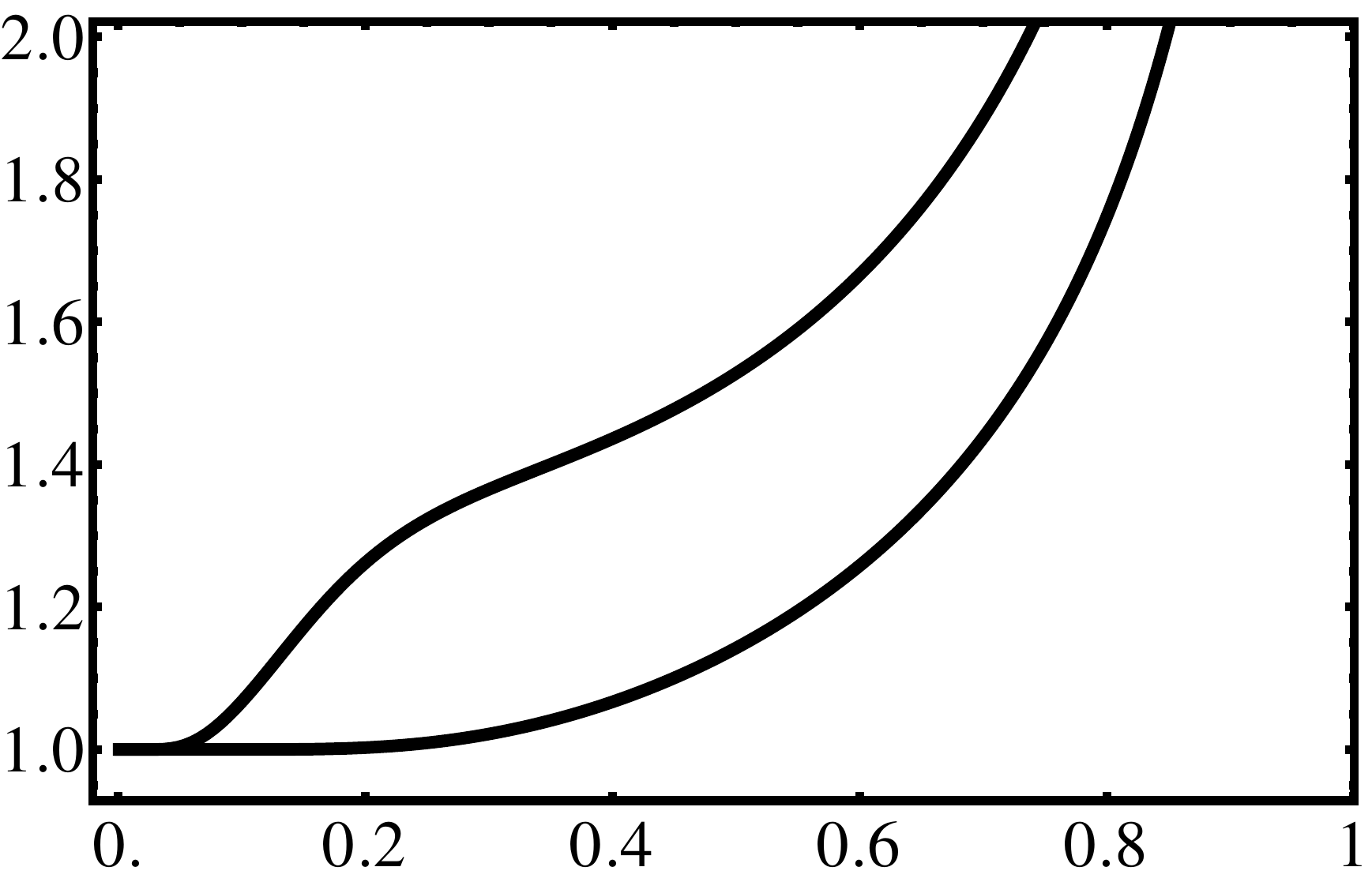} 
	\put(50,-7){\Large $T/\Tc$}
	\put(-5,22){\Large \begin{rotate}{90} $\lambda_{\rm L}(T)/\lambda_{\rm L}(0)$\end{rotate}}
	\put(50,40){\large $(i)$}
	\put(50,20){\large $(ii)$}
	\end{overpic}
\captionsetup{width=.85\textwidth,font=small,labelfont=bf, skip=30pt}
\caption{Temperature dependence of the normalized magnetic penetration depth $\lambda_{\rm L}(T)/\lambda_{\rm L}(0)$ from
equation (\ref{eq:MagnPenDepth}) for intra--band coupling constants  $\lambda_{11}=0.12$ and $\lambda_{22}=0.19$
and inter--band coupling constant $(i)\ \lambda_{12}=0.005$ and $(ii)\ \lambda_{12}=0.07$.}
\label{fig:pen_depth}
\end{figure}
\subsection{Specific heat capacity in two--band superconductors}
The energy change $\delta E_{\vk i}^{\rm ext}$ that is related to a local temperature change $\delta T$
in the $i$--th band can be derived as follows:
\begin{eqnarray*}
\frac{E_{\vk i}+\delta E_{\vk i}^{\rm ext}}{\kB T}&=&\frac{E_{\vk i}+\frac{\partial E_{\vk i}}{\partial T}\delta T}{\kB\left(T+\delta T\right)}
=\frac{1}{\kB T}\left\{E_{\vk i}-\left(\frac{E_{\vk i}}{T}-\frac{\partial E_{\vk i}}{\partial T}\right)\delta T\right\} \\
\delta E_{\vk i}^{\rm ext}&=&-\left(\frac{E_{\vk i}}{T}-\frac{\partial E_{\vk i}}{\partial T}\right)\delta T
\end{eqnarray*}
The change of the BVQP entropy density $\delta\sigma$ due to the temperature change $\delta T$ can be written as
\begin{eqnarray*}
T\delta\sigma&=&\frac{1}{V}\sum_{\vp\sigma i}E_{\vp i}\nu\left(E_{\vp i}+\delta E_{\vp i}^{\rm ext}\right) \\
&=&\frac{1}{V}\sum_{\vp\sigma i}E_{\vp i}\left\{\nu(E_{\vp i})+\frac{\partial\nu(E_{\vp i})}{\partial E_{\vp i}}\delta E_{\vp i}^{\rm ext}\right\} \\
&=&\frac{1}{V}\sum_{\vp\sigma i}y_{\vp i}\left(\frac{E_{\vp i}^2}{T}-E_{\vp i}\frac{\partial E_{\vp i}}{\partial T}\right)\delta T
\end{eqnarray*}
The local thermodynamic relation
\begin{eqnarray*}
T\delta\sigma&=&C_V(T)\delta T
\end{eqnarray*}
allows for the identification of the specific heat capacity $C_V(T)$ of a two--band superconductor in the form
\begin{eqnarray}
\label{eq:SpHeatCapacity}
\begin{aligned}
C_V(T) &\ =\  \sum_{i=1}^2C_{Vi}(T) \\
C_{Vi}(T)&\ =\ \sum_{\vp\sigma}y_{\vp i}\left(\frac{E_{\vp i}^2}{T}-E_{\vp i}\frac{\partial E_{\vp i}}{\partial T}\right)
\end{aligned}
\end{eqnarray}
In Fig.~\ref{fig:sp_heat} we have plotted the dependence of the normalized specific heat capacity $C_{\rm V}(T)/C_{\rm V}(\Tc^+)$ 
on reduced temperature $T/\Tc$ for the set of intra--band pair coupling constants $\lambda_{11}=0.12$ and $\lambda_{22}=0.19$ 
and two values for $\lambda_{12}=0.005\ (i)$ and $\lambda_{12}=0.07\ (ii)$. For the lower value of $\lambda_{12}$ 
the specific heat capacity  shows non--monotonic behavior, which can be explained by thermal activation processes associated with 
the smaller gap $\Delta_1(T)$. Note that the discontinuity of $C_{\rm V}(T)/C_{\rm V}(\Tc^+)$ at the transition temperature depends on $\lambda_{12}$ in a way, 
which will be investigated analytically in more detail in the following section~\ref{sec:SpHeatDiscontinuity}.  
%
%
\begin{figure}[!htb]\ContinuedFloat*
\centering
	\begin{overpic}[width=0.6\textwidth]{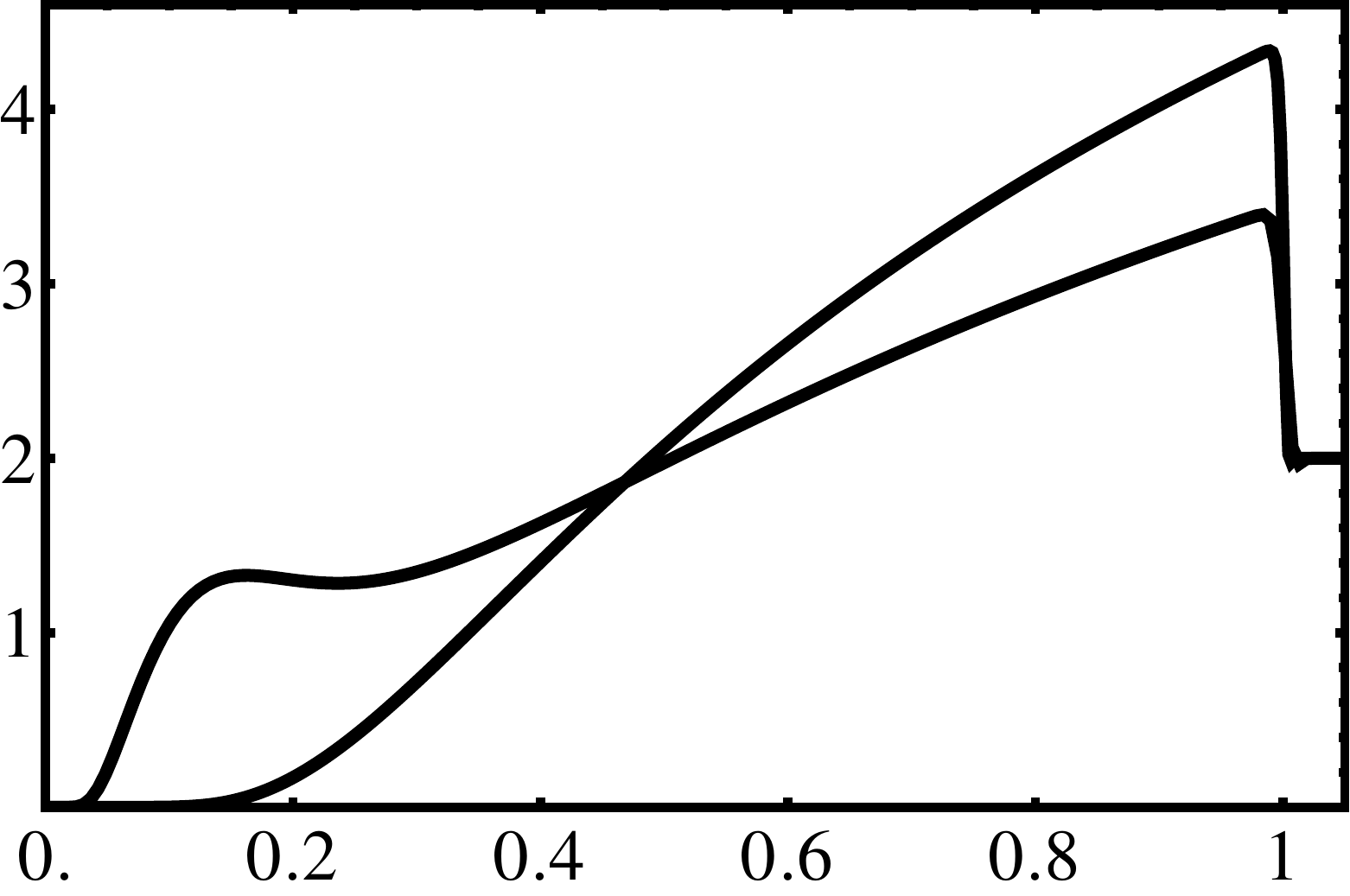} 
	\put(50,-7){\Large $T/\Tc$}
	\put(-5,20){\Large \begin{rotate}{90} $C_{\rm V}(T)/C_{\rm V}(\Tc^{+})$\end{rotate}}
	\put(20,26){\large $(i)$}
	\put(20,13){\large $(ii)$}
	\end{overpic}
\captionsetup{width=.85\textwidth,font=small,labelfont=bf, skip=30pt}
\caption{Temperature dependence of the normalized specific heat capacity $C_V(T)/C_V(\Tc^+)$ from equation (\ref{eq:SpHeatCapacity})
for intra--band coupling constants  $\lambda_{11}=0.12$ and $\lambda_{22}=0.19$ and inter--band coupling constant
$(i)\ \lambda_{12}=0.005$ and $(ii)\ \lambda_{12}=0.07$.}
\label{fig:sp_heat}
\end{figure}
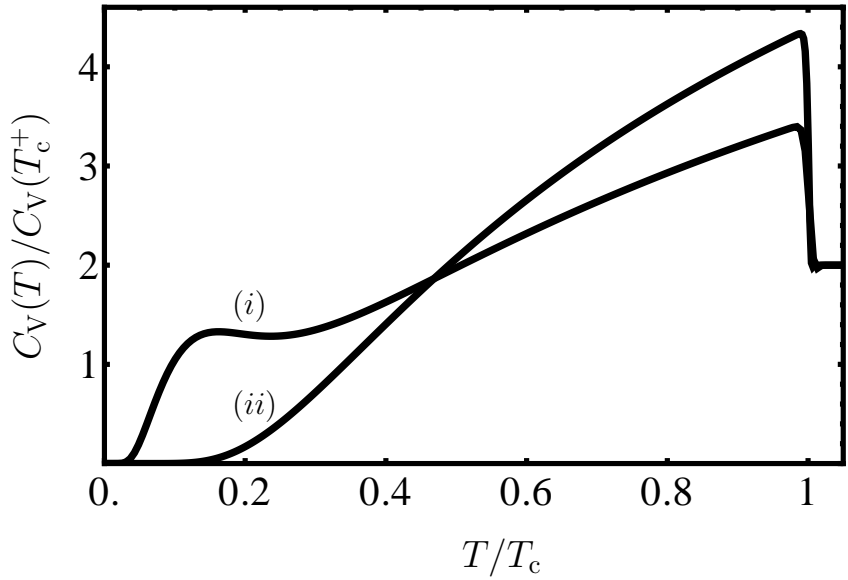
\subsubsection{The specific heat discontinuity in two--band superconductors}
\label{sec:SpHeatDiscontinuity}
In this section we wish to calculate analytically the discontinuity in the specific heat at the transition temperature
and start from the equation (\ref{eq:SpHeatCapacity}):
\begin{eqnarray}
\label{eq:SpHeatCapacityDef}
C_V(T) & = & 2\sum_{\vp i}y_{\vp i}\left[\frac{E_{\vp i}^2}{T}-\frac{1}{2}\frac{\partial\Delta_i^2(T)}{\partial T}\right] \nn \\
           & \stackrel{T\to\Tc}{=}&\sum_{i=1}^2\left\{\frac{N_{i}(0)}{3}\frac{(\pi\kB\Tc)^2}{\Tc}-
             \frac{N_{i}(0)}{2}\lim_{T\to\Tc}\frac{\partial\Delta_i^2(T)}{\partial T}\right\} \\
       & = & \frac{\NF}{3}\frac{(\pi\kB\Tc)^2}{\Tc}-\frac{\NF}{4}\lim_{T\to\Tc}\sum_{i=1}^2\frac{\partial\Delta_i^2(T)}{\partial T} \nn
\end{eqnarray}
In order to make the following calculations more transparent, we have assumed in the latter equality (\ref{eq:SpHeatCapacityDef}), 
that $N_{1}(0)=N_{2}(0)=\NF/2$. The general form of the specific heat discontinuity will be given, however, at the end of this section. 
As a next step we compute the temperature derivatives of the two gap functions $\Delta_1(T)$ and $\Delta_2(T)$ as given
by equation (\ref{eq:EquilGapEq_GL}) near $\Tc$:
\begin{eqnarray*}
\lim_{T\to\Tc}\frac{\partial\Delta_1^2(T)}{\partial T}&=&\frac{8}{7\zeta(3)}\frac{(\pi\kB\Tc)^2}{\Tc}\left[-1+\gamma_{\rm L}\Tc\lim_{T\to\Tc}\frac{\partial r(T)}{\partial T}\right] \\
\lim_{T\to\Tc}\frac{\partial\Delta_2^2(T)}{\partial T}&=&\frac{8}{7\zeta(3)}\frac{(\pi\kB\Tc)^2}{\Tc}\left[-1+\gamma_{\rm L}\Tc\lim_{T\to\Tc}\frac{\partial t(T)}{\partial T}\right]
\end{eqnarray*}
The temperature derivatives of the gap ratios $r(T)$ and $t(T)$ at the transition temperature are obtained from equation (\ref{eq:GapEqSolr_1}) and 
 can be expressed as follows by using the relation $r(\Tc)=1/t(\Tc)$:
\begin{eqnarray*}
\gamma_{\rm L}\Tc\lim_{T\to\Tc}\frac{\partial r(T)}{\partial T}&=&\frac{1-t^2(\Tc)}{1+t^4(\Tc)} \\
\gamma_{\rm L}\Tc\lim_{T\to\Tc}\frac{\partial t(T)}{\partial T}&=&\frac{t^2(\Tc)[t^2(\Tc)-1]}{1+t^4(\Tc)}
\end{eqnarray*}
From this we immediately get
\begin{eqnarray*}
\lim_{T\to\Tc}\frac{\partial\Delta_1^2(T)}{\partial T}
&=& -\frac{8}{7\zeta(3)}\frac{(\pi\kB\Tc)^2}{\Tc}\left[1-\frac{1-t^2(\Tc)}{1+t^4(\Tc)}\right] \\
\lim_{T\to\Tc}\frac{\partial\Delta_2^2(T)}{\partial T}
&=&-\frac{8}{7\zeta(3)}\frac{(\pi\kB\Tc)^2}{\Tc}\left[1+\frac{t^2(\Tc)[1-t^2(\Tc)]}{1+t^4(\Tc)}\right]
\end{eqnarray*}
This can finally be inserted into the expression for $C_V(\Tc^-)$ which leads to
\begin{eqnarray*}
C_V(\Tc^-)&=&\underbrace{\frac{\NF}{3}\frac{(\pi\kB\Tc)^2}{\Tc}}_{=C_V(\Tc^+)}-\frac{\NF}{4}\lim_{T\to\Tc}\sum_{i=1}^2
\frac{\partial\Delta_i(T)}{\partial T} \\
&=&C_V(\Tc^+)\left\{1+\frac{12}{7\zeta(3)}\left[1-\frac{1}{2}\frac{[1-t^2(\Tc)]^2}{1+t^4(\Tc)}\right]\right\}
\end{eqnarray*}
The analytic result for the specific heat discontinuity of two--band superconductors can now be identified to read:
\begin{eqnarray}
\label{eq:SpHeatDiscontinuity}
\begin{aligned}
\frac{\Delta C}{C_N}\equiv\frac{C_V(\Tc^-)-C_V(\Tc^+)}{C_V(\Tc^+)}&\ =\ \underbrace{\frac{12}{7\zeta(3)}}_{=
\left(\frac{\Delta C}{C_N}\right)_{\rm BCS}}\left[1-\frac{1}{2}\frac{[1-t^2(\Tc)]^2}{1+t^4(\Tc)}\right] \\
&\ =\ \left(\frac{\Delta C}{C_N}\right)_{\rm BCS}\left[1-\frac{1}{2}\frac{[1-r^2(\Tc)]^2}{1+r^4(\Tc)}\right]
\end{aligned}
\end{eqnarray}
Accordingly we obtain a modification of the well--known BCS result by the term, which
 depends on the strength of the pairing interaction.\\
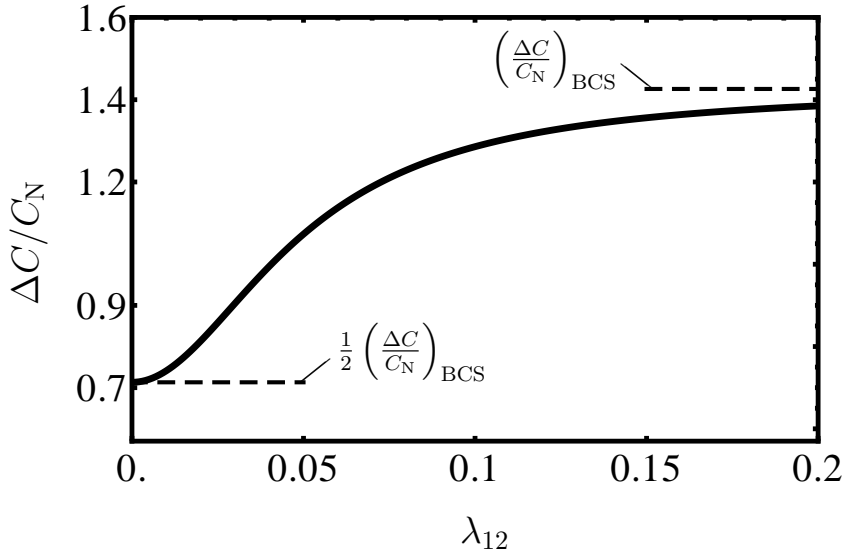
\begin{figure}[!htb]\ContinuedFloat
\centering
	\begin{overpic}[width=0.6\textwidth]{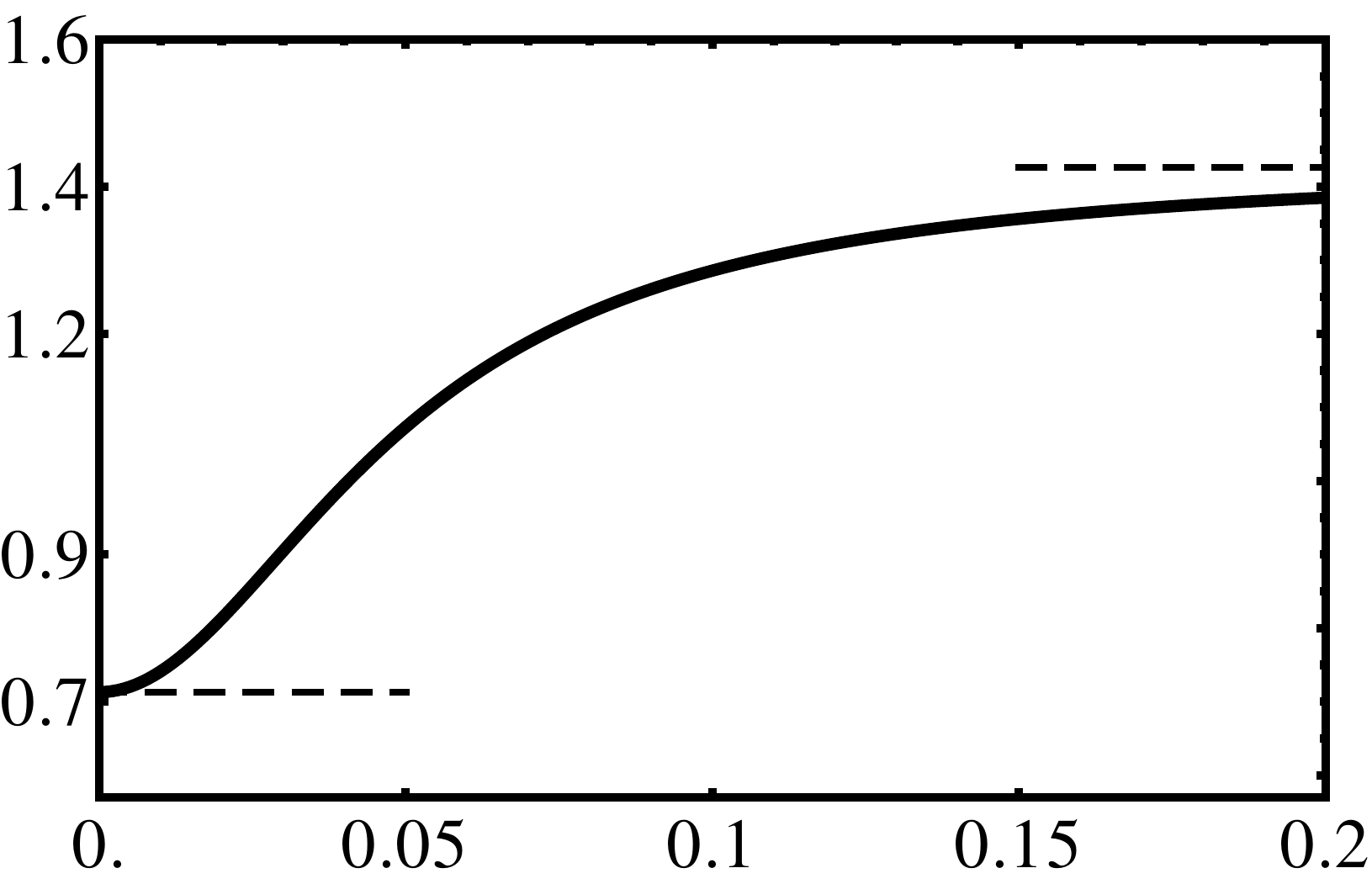} 
	\put(50,-7){\Large $\lambda_{12}$}
	\put(-5,24){\Large \begin{rotate}{90} $\Delta C/C_{\rm N}$\end{rotate}}
	\put(71,55.5){\line(5,-4){4}}
	\put(54,55){\large $\left(\frac{\Delta C}{C_{\rm N}}\right)_{\rm BCS}$}
	\put(33,17){\line(-5,-4){4}}	
	\put(34,17){\large $\frac{1}{2}\left(\frac{\Delta C}{C_{\rm N}}\right)_{\rm BCS}$}
	\end{overpic}
\captionsetup{width=.85\textwidth,font=small,labelfont=bf, skip=30pt}
\caption{Specific heat discontinuity $\Delta C/C_{\rm N}$ as a function of inter--band coupling constant $\lambda_{12}$ 
for intra--band coupling constants $\lambda_{11}=0.12$ and $\lambda_{22}=0.19$.}
\label{fig:sp_heat_discontinuity}
\end{figure}\\
 In order to study the expression (\ref{eq:SpHeatDiscontinuity}), 
 we have plotted the dependence of the specific heat discontinuity 
 on the inter--band coupling constant $\lambda_{12}$ for a two--band superconductor in Fig.~\ref{fig:sp_heat_discontinuity}. 
 As one can see in the figure, the specific heat discontinuity comes close to the BCS--value in the case of large enough $\lambda_{12}$.
 However, in the opposite limit  $(\lambda_{12}\to0)$ it is equal to half of the BCS--value, as can be seen from equation (\ref{eq:SpHeatDiscontinuity}).
 This behavior can been  explained by the fact, that in the latter case only excitations over the large gap give non--vanishing contributions to 
 the specific heat capacity (compare Figs.~\ref{fig:gaps_1912005} and~\ref{fig:gaps_191207}).\\
 \vspace{0.1mm}\\
Now we would like to give the more general
form of the result (\ref{eq:SpHeatDiscontinuity}) for the specific heat discontinuity, which depends on two different 
densities of states $N_{i}(0)$ for each band $i=1,2$, respectively:
\begin{eqnarray*}
\frac{\Delta C}{C_N}&=&\left(\frac{\Delta C}{C_N}\right)_{\rm BCS}
\left\{1-\left[\frac{1-t^2(\Tc)}{1+t^4(\Tc)}\right]\left[\frac{N_{1}(0)}{\NF} - \frac{N_{2}(0)}{\NF}t^2(\Tc)\right]\right\}
\end{eqnarray*}
As a consequence, the specific heat discontinuity of two--band (two--gap) superconductors is seen to
represent the second non--universal BCS--M\"uhlschlegel parameter.  \\
%
\subsection{Spin susceptibility}
\begin{figure}[!htb]
\centering
	\begin{overpic}[width=0.6\textwidth]{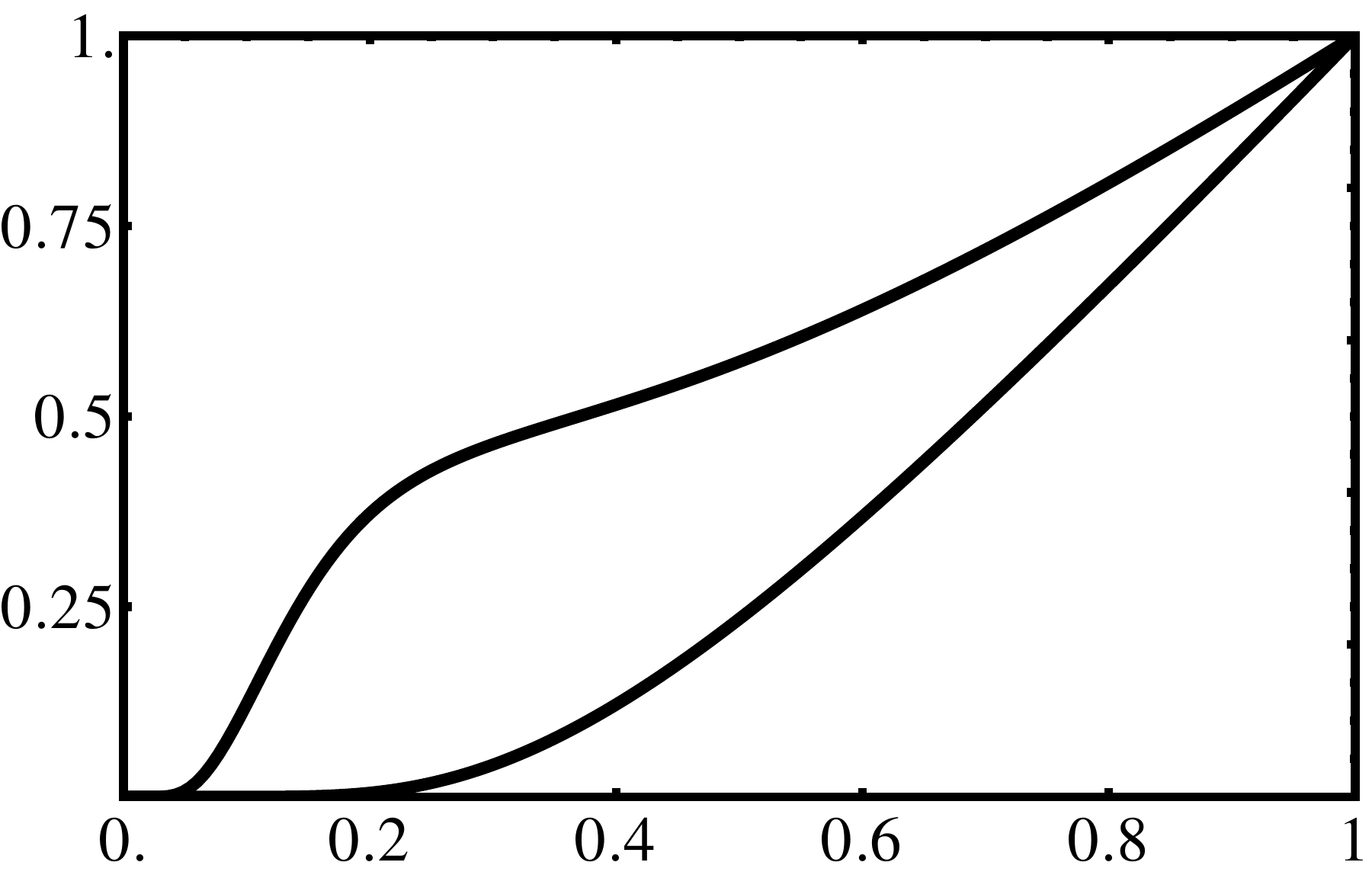} 
	\put(50,-7){\Large $T/\Tc$}
	\put(-5,20){\Large \begin{rotate}{90}$\chi_s(T)/\chi_s(\Tc)$\end{rotate}}
	\put(50,40){\large $(i)$}
	\put(50,22){\large $(ii)$}
	\end{overpic}
\captionsetup{width=.85\textwidth,font=small,labelfont=bf, skip=30pt}
\caption{Temperature dependence of the normalized spin susceptibility $\chi_s(T)/\chi_s(\Tc)$ from equation (\ref{eq:SpinSuscept})
for intra--band coupling constants  $\lambda_{11}=0.12$ and $\lambda_{22}=0.19$ and inter--band coupling constant
$(i)\ \lambda_{12}=0.005$ and $(ii)\ \lambda_{12}=0.07$.}
\label{fig:spin_suscept}
\end{figure}

The spin susceptibility describes the response of the quasiparticle magnetization $\vM$ to an external magnetic field $\vH^{\rm ext}$:
\begin{eqnarray*}
\vM&=&\frac{\gamma\hbar}{2}\sum_{\vp i}{\rm tr}\left\{\vtau\delta\nu\right\} \\
&=&\frac{\gamma\hbar}{2}\sum_{\vp i}{\rm tr}\left\{\vtau\left[\nu(E_{\vp i})
-\frac{\partial\nu(E_{\vp i})}{\partial E_{\vp i}}\frac{\gamma\hbar}{2}\vtau\cdot\vH^{\rm ext}\right]\right\} \\
&=&\left(\frac{\gamma\hbar}{2}\right)^22\sum_{\vp i}y_{\vp i}\ \vH^{\rm ext}=\left(\frac{\gamma\hbar}{2}\right)^2\sum\limits_{i=1}^{2}Y_{i}(T)\ \vH^{\rm ext}=\chi_{\rm s}(T)\vH^{\rm ext}
\end{eqnarray*}
Therefore one may identify the quasiparticle spin susceptibility as 
\begin{eqnarray}
\label{eq:SpinSuscept}
\begin{aligned}
\chi_{\rm s}(T)&\ =\ \sum_{i=1}^2\chi_{{\rm s}i}(T) \\
\chi_{{\rm s}i}(T)&\ \equiv\ \left(\frac{\gamma\hbar}{2}\right)^2Y_{i}(T)
\end{aligned}
\end{eqnarray}
In Fig.~\ref{fig:spin_suscept} we have plotted the dependence of the normalized spin susceptibility $\chi_s(T)/\chi_s(\Tc)$ 
on reduced temperature $T/\Tc$ for the set of intra--band pair coupling constants $\lambda_{11}=0.12$ and $\lambda_{22}=0.19$ 
and two values for $\lambda_{12}=0.005\ (i)$ and $\lambda_{12}=0.07\ (ii)$. For the lower value of $\lambda_{12}$ the spin susceptibility shows 
non--monotonic behavior, which can be traced back again to thermal activation processes associated with the smaller gap $\Delta_1(T)$, as
it has been obtained in the previous local response functions: specific heat capacity and magnetic penetration depth.\\
\vspace{0.1mm}\\
Note that in the case of singlet $s$--wave pairing, the temperature dependence of both the spin susceptibility from Eq.(\ref{eq:SpinSuscept}) 
and the normal fluid density from Eq.(\ref{eq:NormFlDens}) is characterized by the same form of the band--selected Yosida functions $Y_i(T)$.
%
%
\newpage
\section{Summary and conclusions}
\label{sec:SumCon}
This paper is devoted to a comprehensive study of two--band (two--gap) superconductors with spin singlet $s$--wave pairing.
The treatment of these systems requires a generalization of the microscopic BCS description of global equilibrium
and weak response to the case that two bands, on which the gaps $\Delta_i(T), i=1,2$ reside, cross the Fermi surface.
From this generalized BCS description there emerge a couple of remarkable aspects, which deserve being spotlighted in what follows:
\begin{itemize}
\itemsep 0pt
\parskip 0pt
\item Two--band superconductors are seen to display a common transition temperature only in the presence of a finite
inter--band pair--coupling constant $\lambda_{12}$.
\item In one--band superconductors there are two so--called BCS--M\"uhlschlegel~\cite{MUEHLSCHLEGEL1959} parameters, namely
the gap at zero temperature $\Delta(0)/\pi\kB\Tc$ and the specific heat discontinuity $\Delta C/C_{\rm N}$ at the transition
\begin{eqnarray*}
\left[\frac{\Delta(0)}{\pi\kB\Tc}\right]_{\rm BCS}=\frac{1}{{\rm e}^{\gamma}}\ \ ;\ \
\left[\frac{\Delta C}{C_{\rm N}}\right]_{\rm BCS}=\frac{12}{7\zeta(3)}
\end{eqnarray*}
which are universal in the sense that they may depend on the nodal structure implied by the possible unconventionality
of the pairing~\cite{EINZEL2002}, but they do not depend on the characteristic frequency $\epsilon_0/\hbar$ of the exchange boson
and the strength $\lambda=N(0)\Gamma$ of the pairing interaction. In two--band superconductors with small but finite
inter--band coupling $\lambda_{12}$ this universality gets lost and both $\Delta_i(0)/\pi\kB\Tc, i=1,2$ and $\Delta C/C_{\rm N}$,
which can both be evaluated analytically, are seen to depend on $\lambda_{12}$:
\begin{eqnarray*}
\frac{\Delta_1(0)}{\pi\kB\Tc}&=&\left[\frac{\Delta(0)}{\pi\kB\Tc}\right]_{\rm BCS}{\rm e}^{-\gamma_{\rm L}[r(\Tc)-r(0)]} \ \ ; \ \ 
\frac{\Delta_2(0)}{\pi\kB\Tc}=\left[\frac{\Delta(0)}{\pi\kB\Tc}\right]_{\rm BCS}{\rm e}^{-\gamma_{\rm L}[t(\Tc)-t(0)]}\\
\frac{\Delta C}{C_{\rm N}}&=&\left[\frac{\Delta C}{C_{\rm N}}\right]_{\rm BCS}\left\{1-\frac{1}{2}\frac{[1-r^2(\Tc)]^2}{1+r^4(\Tc)}\right\}
=\left[\frac{\Delta C}{C_{\rm N}}\right]_{\rm BCS}\left\{1-\frac{1}{2}\frac{[1-t^2(\Tc)]^2}{1+t^4(\Tc)}\right\}
\end{eqnarray*}
with $r(\Tc)=1/t(\Tc)$ from (\ref{eq:rtGL}) and $r(0)=1/t(0)$ from (\ref{eq:rtLT}).
\item While the temperature dependence of the energy gap in one--band superconductors is known to be strictly monotonic,
the smaller of the gaps in two--band superconductors displays non--monotonic behavior in the limit where $\lambda_{12}$
gets sufficiently small.
\item The knowledge of the full temperature dependence of the two gaps $\Delta_i(T)$ is a necessary prerequisite
for the calculation of the relevant local response functions of the superconductor under consideration.
In one--band superconductors these local response functions, namely the normal fluid density, the magnetic
penetration depth, the specific heat capacity and the spin susceptibility, are known to decrease monotonically with decreasing
temperature and display thermally activated behavior in the low temperature limit. In contrast, the local response functions
of two--band superconductors may show a non--monotonic decrease with decreasing temperature particulary in the case
of very small inter--band pair--coupling constants $\lambda_{12}$. In such a case all these response functions display a hump
at the low temperature end, which can be associated with the activated behavior connected with the smaller gap.
\end{itemize}
All the results obtained in this paper form the basis for a general discussion of the dynamic response of two--gap
superconductors. The dynamics of the {\it phase} of the order parameter will be characterized, besides the Nambu--Goldstone
boson~\cite{NAMBU2009} or gauge mode, by a new massive collective mode, which was first discussed in the literature by A. J. Leggett ~\cite{LEGGETT1966}.
This so--called Leggett--mode owes its very existence to the finiteness of the Leggett--parameter $\gamma_{\rm L}=\lambda_{12}/{\rm det}\vlambda$
(\ref{eq:LeggettParam}) and is seen to scale with the product of the two gaps $\Delta_1(T)$ and $\Delta_2(T)$, which we have calculated rigorously
in section 2 of this paper. The frequency of the Leggett--mode reads~\cite{LEGGETT1966}
\begin{eqnarray*}
\hbar^2\omega^2=4\gamma_{\rm L}\Delta_1(T)\Delta_2(T)\frac{\lela\lambda\rira_1+\lela\lambda\rira_2}{\lela\lambda\rira_1\lela\lambda\rira_2}
\end{eqnarray*}
with $\lela\lambda\rira_i, i=1,2$ the dimensionless condensate densities on the two bands. The Leggett--mode
turns out to be unaffected by the long--range Coulomb interaction and therefore by the Higgs mechanism. Its existence
leads to numerous consequences for the dynamic response of two--band superconductors, which we plan to publish in two forthcoming papers,
namely on the electromagnetic response~\cite{BANDE-2-2013} and on the electronic Raman response~\cite{BANDE-3-2013}.
\section*{Acknowledgement}
The authors are grateful to W. Biberacher, B. S. Chandrasekhar, R. Gross, R. Hackl, M. Kartsovnik, L. Klam and D. Manske for enlightening discussions.
\footnotesize
\itemsep 0pt
\parskip 0pt
\bibliographystyle{unsrt}
\bibliography{biblio}
\end{document}